\documentclass[fleqn,usenatbib]{mnras}

\usepackage{newtxtext,newtxmath}

\usepackage[T1]{fontenc}

\DeclareRobustCommand{\VAN}[3]{#2}
\let\VANthebibliography\thebibliography
\def\thebibliography{\DeclareRobustCommand{\VAN}[3]{##3}\VANthebibliography}

\usepackage{graphicx}	
\usepackage{amsmath}	
\usepackage{gensymb}    

\usepackage{xcolor}
\usepackage{verbatim}
\usepackage{orcidlink}

\usepackage{svg} 

\usepackage{tabularx} 
\usepackage{blindtext} 
\usepackage{color, colortbl}
\usepackage{adjustbox, booktabs}
\usepackage{array, caption}

\newcommand{\Kepler}{\emph{Kepler}}
\newcommand{\TESS}{\emph{TESS}}
\newcommand{\Gaia}{\emph{Gaia}}
\newcommand{\tess}{\emph{TESS}}	
\newcommand\rsun{\ensuremath{R_\odot}}		
\newcommand\rearth{\ensuremath{R_\oplus}}	
\newcommand{\deathstar}{\texttt{DEATHSTAR}}
\newcommand{\mycomment}[1]{}

\definecolor{new_color}{HTML}{CF0000} 

\definecolor{html_black}{HTML}{000000}

\newcommand\bedit[1]{\textcolor{html_black}{#1}} 
\newcommand\bedittwo[1]{\textcolor{html_black}{#1}}
\usepackage{lineno}

\definecolor{table_highlight}{HTML}{0dde95} 
\definecolor{table_alternate}{HTML}{f0f0f0} 
\renewcommand{\tabcolsep}{1pt} 
\newcolumntype{+}{>{\global\let\currentrowstyle\relax}}
\newcolumntype{^}{>{\currentrowstyle}}

\title[TESS ``follow-up'' with archival ground-based data]{\deathstar: A system for confirming planets and identifying false positive signals in \TESS\ data using ground-based time domain surveys}

\author[Ross et al.]{Gabrielle Ross \orcidlink{0009-0006-7023-1199},$^{1,2,3}$\thanks{E-mail: gr8740@princeton.edu}
Andrew Vanderburg \orcidlink{0000-0001-7246-5438},$^{2}$
Zo{\"e}\ \bedit{L.} de Beurs \orcidlink{0000-0002-7564-6047},$^{4,5}$
Karen A. Collins \orcidlink{0000-0001-6588-9574},$^{6}$
Rob J. Siverd \orcidlink{0000-0001-5016-3359},$^{7}$\newauthor
Kevin Burdge \orcidlink{0000-0002-7226-836X},$^{2,8}$
\\
$^{1}$Princeton University, Princeton, NJ 08544\\
$^{2}$Department of Physics and Kavli Institute for Astrophysics and Space Research, Massachusetts Institute of Technology, Cambridge, MA 02139, USA\\
$^{3}$The Brearley School, 610 E 83rd St, New York, NY 10028\\
$^{4}$Department of Earth, Atmospheric and Planetary Sciences, Massachusetts Institute of Technology,  Cambridge,  MA 02139, USA\\
$^{5}$NSF Graduate Research Fellow, MIT Presidential Fellow, MIT Collamore-Rogers Fellow\\
$^{6}$Center for Astrophysics $|$ Harvard and Smithsonian, 60 Garden Street, Cambridge, MA 02138, USA\\
$^{7}$Institute for Astronomy, University of Hawaii at Manoa, Honolulu, HI 96822, USA\\
$^{8}$Pappalardo Fellow\\
}

\date{Accepted XXX. Received YYY; in original form ZZZ}

\pubyear{2022}

\begin{document}
\label{firstpage}
\pagerange{\pageref{firstpage}--\pageref{lastpage}}
\maketitle

\begin{abstract}
We present a technique for verifying or refuting exoplanet candidates from the \textit{Transiting Exoplanet Survey Satellite} (\TESS) mission by searching for nearby eclipsing binary stars using higher-resolution archival images from ground-based telescopes. Our new system is called \textit{Detecting and Evaluating A Transit: finding its Hidden Source in Time-domain Archival Records} (\deathstar). We downloaded time series of cutout images from two ground-based telescope surveys (the Zwicky Transient Facility, or ZTF, and the Asteroid Terrestrial-impact Last Alert System, or ATLAS), analyzed the images to create apertures and measure the brightness of each star in the field, and plotted the resulting light curves using custom  routines. Thus far, we have confirmed on-target transits for $17$ planet candidates, and identified $35$ false positives and located their actual transit sources. With future improvements to automation, \deathstar\ will be scaleable to run on the majority of TOIs.
\end{abstract}

\begin{keywords}
eclipses, exoplanets, (stars:) binaries: eclipsing
\end{keywords}



\section{Introduction} \label{sec:intro}
All astronomical surveys designed to detect exoplanets also generate false positive signals that can closely mimic exoplanets and which must be separated from the real planet candidates before they can be confirmed. In many cases, these false positives dramatically outnumber the signals of real exoplanets. Despite the fact that the first real exoplanet signals were detected only about 30 years ago \bedittwo{\citep{campbellwalker, 1995Natur.378..355M}}, false positives have been around for over a century \citep{jacob}. 

Identifying false positives is particularly important for exoplanet surveys using the transit method. This is because a fairly common astrophysical situation (a faint or distant eclipsing binary star blended with a brighter star) can cause highly realistic approximations of transiting exoplanets \citep{brown2003, charbs2004}. Indeed, very early transit surveys like Project Vulcan \citep{borucki2001} had high false positive rates, with dozens of candidates sent for spectroscopic followup \citep{lathamvulcan} but no planet discoveries. Even after the breakthrough discovery of transiting planets in the Optical Gravitational Lensing Experiment (OGLE) survey \citep{ogletr56, konacki1}, false positives from blended eclipsing binaries remained a concern \citep{Torres2004}. Nevertheless, over the next decade, surveys like the Transatlantic Exoplanet Survey (TrES, \citealt{alonso2004}), the Hungarian Automated Telescope Network exoplanet survey (HATNet, \citealt{bakos2004}), the Wide Angle Search for Planets (WASP, \citealt{Pollacco:2006}), and the Kilodegree Extremely Little Telescope (KELT, \citealt{pepper2007}) developed procedures and tools \citep{cc, lathamhatnet, collins2018} to identify and rule out false positives that made transiting exoplanet discovery routine.  

The launch of the \Kepler\ space telescope \bedittwo{\citep{2010Sci...327..977B, 2010ApJ...713L..79K}} in 2009 revolutionized the field of transiting exoplanet discovery. Suddenly, the transiting exoplanet community was presented data with photometric precision orders of magnitude better than ground-based transit surveys. The first planet detections from \Kepler\ were only confirmed after extensive observations to rule out false positives, following the best-practices learned in the previous decade of ground-based discovery \citep{borucki2010}. But the community quickly realized that many of these precautions were no longer necessary in the \Kepler\ era. Kepler had several advantages over the ground-based surveys of the previous decade:
\begin{enumerate}
    \item Extremely high photometric precision, which makes it easier to identify subtle features that signify false positives \citep[e.g.][]{faigler}.
    \item Sensitivity to smaller transiting planets, which are intrinsically more common and therefore less likely to be false positives \citep{mortonjohnson2011}. 
    \item Sensitivity to multiplanet systems, which are much less likely to be mimicked by false positives \citep{latham2011, lissauer, rowe}.
    \item A \bedit{better angular resolution} (4\arcsec\ pixels and 3.1-7.5\arcsec\ FWHM, \citealt{koch}) than most (but not all) previous surveys, which decreases the likelihood of blended light from eclipsing binary stars. 
\end{enumerate}
Overall, these advantages led to a low (less than 10\%) false positive probability for most \Kepler\ planet candidates \citep{mortonjohnson2011}, dramatically reduced the scrutiny needed to confidently consider any given signal as a validated exoplanet, and enabled the discovery of thousands of new exoplanets that have revolutionized the field \citep{rowe, morton2016}.

\begin{figure*}
\centering
\includegraphics[width=\textwidth]{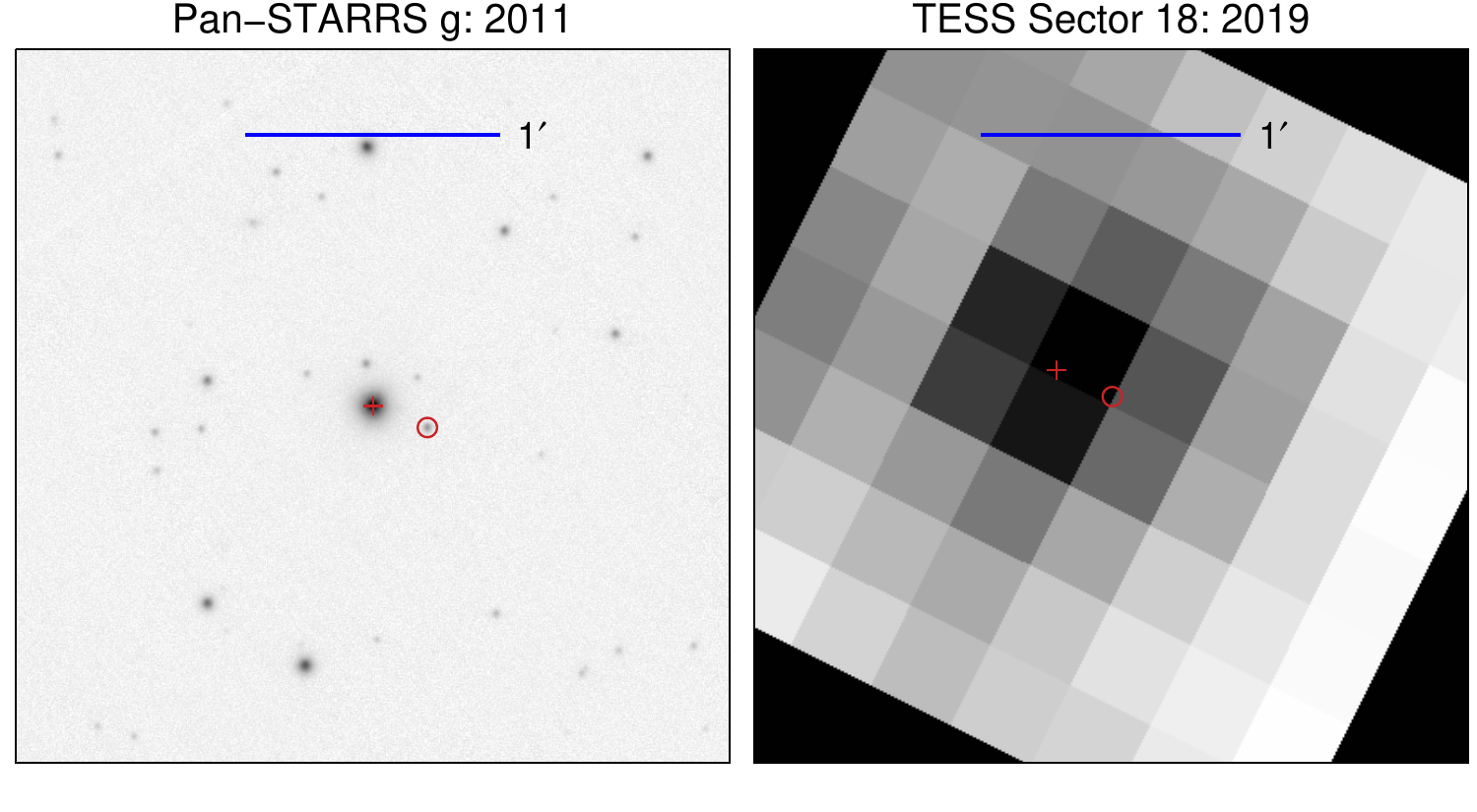}
\caption{Images of the TESS planet candidate TOI 4148.01 that show the difference in spatial resolution between seeing-limited ground-based observations and \TESS\ data. We show a ground-based image from Pan-STARRS in the left panel and an image of the same field in \TESS\ in the right panel. The cross marker shows the location of original target from \TESS\ in both images and the circle shows the location of the actual signal of this TOI. While both of these stars are clearly visible and easily distinguished in the ground data from Pan-STARRS, the majority of the same field is completely blended in \TESS. This blending results in many false positives for \TESS. \label{TESS vs. Pan-STARRS}}
\end{figure*}

In 2018, right as the \Kepler\ spacecraft exhausted its fuel reserves and was retired, its successor mission, the Transiting Exoplanet Survey Satellite \citep[\TESS;][]{ricker} was launched. \TESS\ was designed to be a wide-field version of \Kepler\ and extend its power from relatively small patches to nearly the whole sky. \TESS\ retains many of the same advantages (especially its high photometric precision and sensitivity to small, multiplanet systems) that made \Kepler\ so successful, and it has discovered thousands of new planet candidates in its own right \citep{guerrero, kunimoto2022}. However, in order to cover its wide field of view and efficiently downlink the data to Earth, \TESS\ has significantly larger pixels than the \Kepler\ ($21''$ for \TESS\ versus $4''$ for \Kepler), which increases the likelihood of false positive contamination. Figure \ref{TESS vs. Pan-STARRS} shows the extent of the problem of TESS false positives by comparing images of the same field from  \TESS\ and the ground-based Panoramic Survey Telescope and Rapid Response System \citep[Pan-STARRS;][]{Tonry_2012}. \bedit{One advantage of  observing in space with missions like \TESS\ is that we do not observe through atmospheric turbulence (which allows us to produce clearer light curves compared to ground-based telescopes). However,  \TESS's larger pixel-size (and lower spatial resolution) increases its false-positive rate due to blending. In order to have confidence in the \TESS\ planet candidate list, these potential false positives must be addressed and ground-based telescopes with higher spatial resolution can help us achieve this.}

To combat this increased risk of false positives from stars blended in the large \TESS\ pixels, the \TESS\ team formed the \TESS\ Followup Observation Group (TFOP) Sub-group 1 (SG1), which conducts ground-based light curve follow-up of planet candidates (called \TESS\ Objects of Interest, or TOIs) to determine if the \TESS\ signal is on or off-target. SG1 has observed 3200 of the 5900 planet candidates, using dozens of telescopes on the ground and in space. This process has taken thousands of hours of telescope time and human effort to analyze the new data. Given the importance and expense of this effort, taking advantage of existing datasets to perform similar analysis while minimizing both telescope and human time would be highly beneficial.

In this paper, we introduce \deathstar, a system for confirming false positives using existing ground-based data. \deathstar\ uses data from two publicly available time-domain surveys, the Asteroid Terrestrial-impact Last Alert System (ATLAS; \citealt{2021TNSAN...7....1S, 2018PASP..130f4505T}) and the Zwicky Transient Facility (ZTF; \citealt{mascri, 2019PASP..131a8002B}), to identify false positive contaminants in \TESS\ observations. \deathstar\ is a pipeline that downloads either \bedit{calibrated} ZTF or ATLAS \bedit{pixel} data, analyzes the images by collecting brightnesses for each star to create light curves for each object in the field and plots those light curves folded on the planet candidate's orbit. These subplots\footnote{\bedittwo{We note that we use the word ``subplot'' to describe graphs with their own axes among a greater collection of plots in the same image; here, each subplot shows the light curve of an individual star in the field.}} allow the user to quickly assess if there is a transit dip in any of the nearby stars other than the target, which would indicate a false positive. Although these surveys are ground-based and have worse photometric precision than \TESS, they have higher spatial resolution and can thus in many cases identify blended contaminants in \TESS. Our work is inspired by that of \cite{2022arXiv220905845P}, who performed an analogous analysis to ours using time series photometry from the \Gaia\ mission. \Gaia\ has better spatial resolution and photometric precision than the ground-based surveys we consider, but\bedittwo{\ observes each star relatively few times}, so our work with ZTF and ATLAS is highly complementary to this effort.

Our paper is organized as follows. In Section \ref{sec:method}, we describe the methods we used to calculate and plot the light curves. In Section \ref{sec:Results}, we show the results for a specific handful of systems. In Section \ref{sec:discussion}, we discuss the implications of this work for follow-up of \TESS\ planet candidates and our plans for future work, and we conclude in Section \ref{sec:conclusions}.

\section{Methods} \label{sec:method}
\subsection{Downloading data and extracting photometry} \label{sec:analyzing}
This section discusses how we analyzed images from ATLAS and ZTF and measured each star's brightness. In brief, we downloaded the ground-based datasets for our targets, identified the precise location of the stars in each frame, and recorded the total brightness of each star using aperture photometry. In detail, we took the following steps for each TESS planet candidate we analyzed:

\begin{figure*}
\centering
\includegraphics[width=\textwidth]{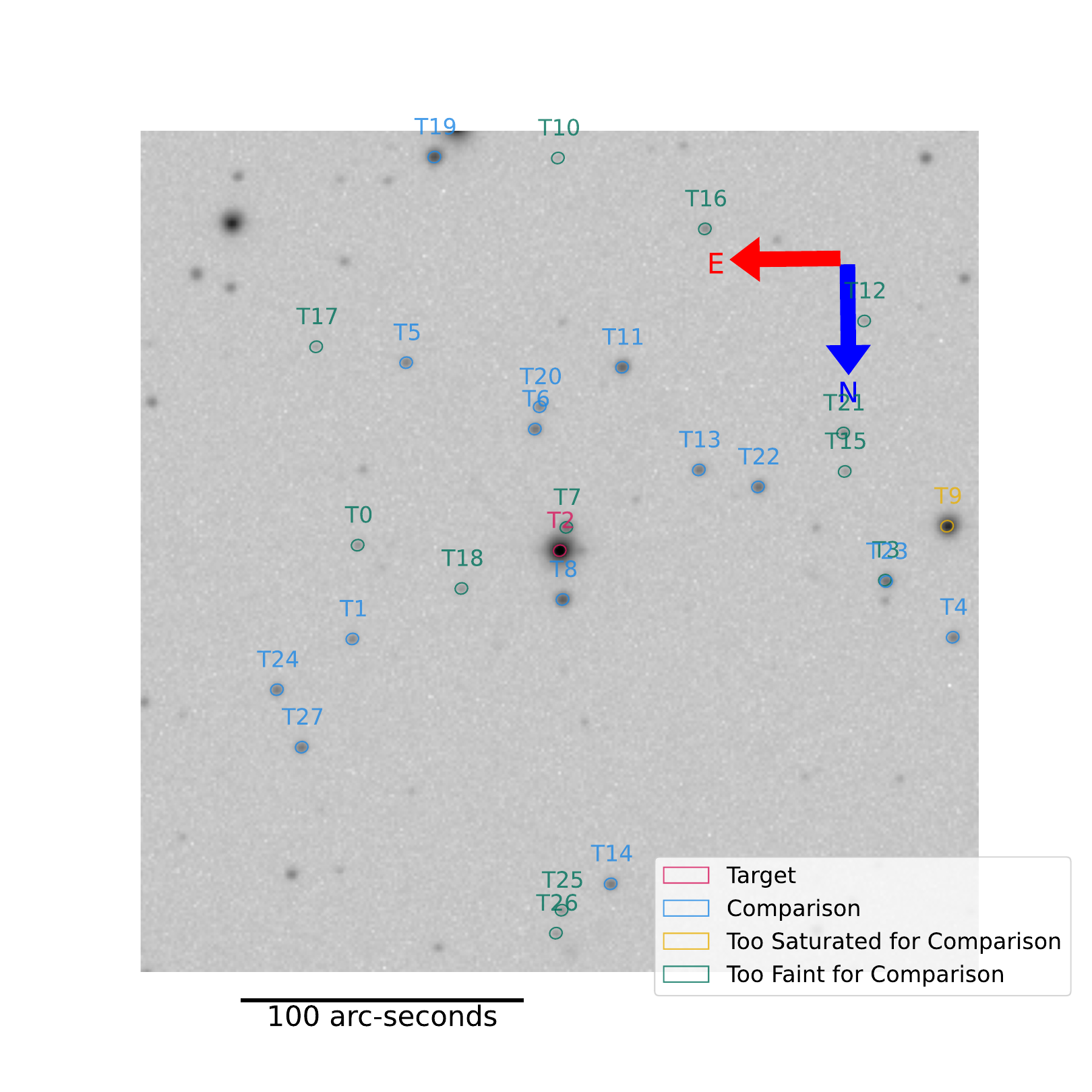}
\caption{A red optical ZTF image of the field surrounding TOI 1423. Different colored ellipses surround stars identified as the target planet candidate host (pink), comparisons used to calculate differential photometry (light blue), saturated stars too bright for use as comparisons (yellow), and stars too faint for use as comparisons (green). The plotted ellipses are the actual photometric apertures used for light curve extraction. Even though some stars are too saturated or dim to use as comparison stars, we extract and plot their light curves anyway since they can still be the source of the transit signal. This particular reference image corresponds to one point in the light curves shown in Figure \ref{Light Curve}. \label{reference_image}}
\end{figure*}

\begin{enumerate}
\item We started by identifying all of the stars nearby to our target planet candidates by querying the \TESS\ Input Catalog (TIC, \citealt{stassun}) from the Barbara A. Mikulski Archive for Space Telescopes (MAST). Following the convention of TFOP SG1 observations, we queried information from all the stars within a $2.5$ arc-minute radius of the planet candidate's position. The TIC query gave us a comprehensive list of the nearby stars' properties, among which we were particularly interested in the positions, \TESS\ magnitudes (Tmag), and proper motion\bedittwo{s mostly from \Gaia\ DR2 \citep{2018A&A...616A...1G}}. \bedit{If a given \TESS\ planet candidate is in fact a false positive from a nearby eclipsing binary, then the contaminant is likely one of these nearby stars listed in the TIC.} 
\item We then downloaded cutouts of images from ground-based time-domain surveys showing our target and the neighboring stars we identified in our query to the TIC. We used two ground-based telescope datasets that both have higher image resolution than \TESS\ and many observations of a large fraction of the sky: ZTF \citep{2019PASP..131a8002B} and ATLAS \citep{2018PASP..130f4505T}. ZTF observes the entire Northern sky with $3$ different filters ($g, r, i$) a cadence of about $2$ days from the $48$ inch telescope at Palomar Observatory. ZTF is designed for the discovery of transient events like stellar explosions in other galaxies, so it has a relatively faint saturation magnitude around $12.5-13.2$ in its various filters. Although bright stars, like our planet candidate target stars, often saturate ZTF's images, we chose to use it because it is highly sensitive to stars with \TESS-band magnitudes between Tmag = $13$ and Tmag = $19$. False positive contaminants that mimic planetary transits often come from such dim stars. On the other hand, ATLAS consists of four $0.5$ meter telescopes: two in Hawaii, one in Chile, and one in South Africa. ATLAS was first commissioned in the Northern hemisphere, and has a long ($5$ year) baseline of observations in the North. The Southern telescopes were commissioned more recently, so there are fewer Southern observations (about $1$ year), though data collection is ongoing. \bedit{The ATLAS telescopes survey the sky regularly using one of two filters: an ``orange'' filter with $560-820$ nm bandpass and a ``cyan'' filter with $420-650$ nm bandpass. The orange filter lessens the impact of sky brightness and is used daily. The cyan filter is employed by one telescope per hemisphere exclusively when the moon is set. The ATLAS system uses a 10k x 10k CCD (model STA1600). It has a $5.5$-degree FOV with a $1.86$ arcsecond per pixel scale. The finer pixel scale of ATLAS data allows it to identify blending in TESS data.} ATLAS is designed to detect near-Earth asteroids and prioritizes observing cadence over survey depth, collecting multiple observations of the entire sky each night. Although it does not go as deep as ZTF, we chose to use it because of its higher cadence. In addition, the ATLAS dataset contains more images from the Southern Celestial Hemisphere, which enables the study of significantly more targets than ZTF alone. In this way, ZTF and ATLAS provide complementary datasets\footnote{\bedittwo{We note ATLAS has its own photometry pipeline, but it only calculates the light curve of one star at a time, rather than all stars in the field simultaneously. It is therefore significantly slower at calculating the light curve of the dozens of stars in typical fields than the pipeline described in this work.}}.

We then retrieved image data of planet candidate host stars from the two surveys. For ZTF, we designed a programmatic query method using the \texttt{ztfquery} module \citep{ztfquery}. We modified the package to allow for retrieving $5\times5$ arcminute cutouts ($300$ pixels by $300$ pixels with $1$ pixel as $1$ arcsecond) around an input RA and DEC instead of the full CCD images ($3000$ pixels by $3000$ pixels), which significantly cut down on download time and storage requirements. For the ATLAS data retrieval, we used the \texttt{Fallingstar} task queue \bedit{\citep{2021TNSAN...7....1S}}\footnote{\url{https://fallingstar-data.com/forcedphot/queue/}} and manually input the RA and DEC to download the $12.4$ arc-minute cutouts ($400$ pixels by $400$ pixels with $1$ pixel as $1.86$ arc-seconds). In order to retrieve all of the ATLAS images, we query the data by using the GUI API multiple times (starting at the time of the last image returned from the previous image grouping), as each submission can only return the first $500$ images. In the future, we plan to automate the acquisition of ATLAS data similar to ZTF.

\item To analyze the retrieved images, we looped through each image to identify the location and measure the brightness of each star in the field (both the target star identified by the \TESS\ project as the planet candidate host and each possible contaminant star within 2.5 arcminutes). We started by using the astrometric solution from the ATLAS and ZTF images to get an initial guess of the location of each star in the field. Because the ATLAS and ZTF datasets span many years, stars with high proper motion can move significantly in the images over the course of the observations. Therefore, we calculated the RA and Dec for each star in the images at the time of each image (converted into a decimal year from Julian Date) using the proper motion information from the TIC. To convert our RA and Dec values into pixel values to compare in our images, we used the \texttt{astropy.wcs} Python package \citep{astropy:2013,astropy:2018,astropy:2022} to convert these time-dependent RA and DEC positions to pixel coordinates in each frame. Although the ZTF images worked ``out of the box'' with the standard \texttt{astropy.wcs} routines, the ATLAS fits files use higher order distortion terms than are natively recognized by that package. Thus,  we made minor edits to the fits headers and use different software routines (\texttt{astropy.wcs.wcs\_world2pix} for ZTF, and \texttt{astropy.wcs.all\_world2pix} for ATLAS).

\item Once we derived an initial estimate of the location of each star, we performed a least-squares fit to a star in the field to both measure the image point spread function's size and shape, as well as more precisely measure the positions of stars in the image (mostly important for ZTF). We manually chose a relatively bright but unsaturated, isolated star and performed a least squares fit to the image surrounding the star with a  two-dimensional elliptical Gaussian profile. We allowed the elliptical Gaussian to have different semimajor and semiminor axes with arbitrary rotation, which helps model stars for which the image point spread function is elongated. In particular, we defined a model for the Gaussian $g(x,y)$ as a function of the $x$ and $y$ pixel positions such that 

\begin{equation}
g(x,y) = m + pe^{- [a(x-x_0)^2 + 2b(x-x_0)(y-y_0) + c(y-y_0)^2]}
\end{equation}
where
\begin{equation}
a = \frac{\cos^2\theta}{2\sigma_x^2} + \frac{\sin^2\theta}{2\sigma_y^2},
\end{equation}
\begin{equation}
b = -\frac{\sin(2\theta)}{4\sigma_x^2} + \frac{\sin(2\theta)}{4\sigma_y^2},
\end{equation}
\begin{equation}
c = \frac{\sin^2\theta}{2\sigma_x^2} + \frac{\cos^2\theta}{2\sigma_y^2},
\end{equation}
and where $\sigma_x$ and $\sigma_y$ are the lengths of the two axes of the ellipse, $m$ is a constant offset representing the background level of the image, $x_0$ and $y_0$ are the $x$ and $y$ central pixel locations of the star, $p$ is the peak brightness of the Gaussian, and $e$ is Euler's number, the base of the natural logarithm. The Gaussian fit was performed using the \texttt{mpfit}\footnote{\raggedright\url{https://github.com/segasai/astrolibpy/blob/master/mpfit/mpfit.py}} \citep{2009ASPC..411..251M} module, which implements the Levenberg-Marquardt least-squares minimization algorithm. We initialized the fit with starting guesses for $x_0$ and $y_0$ from the \texttt{WCS} coordinates, a guess for $m$ from the median pixel value in the image, and a guess for $p$ from the brightest pixel value within a $20$ pixel by $20$ pixel square area surrounding the star. We initialized $\sigma_x$ and $\sigma_y$ with guesses of $3$ pixels, and initialized the rotation angle  $\theta$ at 0 degrees. 

From the fit for each image, we recorded the offset from the predicted pixel location of the star and our best-fit values for $x_0$ and $y_0$, as well as the rotation angle and ellipse axes $\sigma_x$ and $\sigma_y$. We took the fine-tuned $x$ and $y$ pixel offsets and parameters describing the width, elongation, and rotation of the point spread function to be representative of the rest of the stars in the image because they should all be affected by the same observing conditions. We found that generally, the location of the star we measured in our fits was more accurate than the estimates from the ZTF astrometric solutions, so we applied the offset between the two estimates to positions of all stars in the field. We found that this was not necessary for the ATLAS images, so did not perform this step for ATLAS.
\item We then took the results of our fit to the profile of a single isolated star and used them to define photometric apertures to produce light curves for all of the stars in the vicinity of the \TESS\ candidate host star. We centered the apertures at the coordinates of each star in the image predicted by the image's astrometric solution, and for ZTF, modified by the offset we measured in our fit. For ATLAS we do not use this fit for the positioning of the aperture as the \texttt{astropy.wcs.all\_world2pix} does not need an additional offset adjustment. We defined elliptical apertures using \texttt{photutils} \citep{photutils}, an \texttt{astropy} affiliated package for extracting the brightnesses of astronomical sources from images. The rotation and length of the elliptical axes came directly from the elliptical Gaussian fit described above. We experimented with increasing the size of the elliptical apertures by scaling the $\sigma_x$ and $\sigma_y$ values by a scalar, but found optimal results when the scale factor was 1 (that is, the aperture axes were equal to $\sigma_x$ and $\sigma_y$). We show a typical image with apertures overlaid in Figure \ref{reference_image}.
\item We performed background subtraction in our images by subtracting the median pixel value of the image cutouts. We found ATLAS images showed noisier and less uniform backgrounds, so we performed an additional step. We subtracted the mean of the minimum values of each row within the image and then set all the pixels less than 0 equal to 0. \bedittwo{We also experimented with estimating the background flux by measuring the flux in an annulus surrounding each star, but we found no significant improvement from this method.} 
\item With the apertures defined and background subtraction performed, we used \texttt{photutils} to calculate the total brightness inside the ellipses. We also recorded the maximum pixel brightness from the aperture in order to identify saturated observations.
\item Lastly, after performing these steps for each star in each image, we saved the pixel locations and brightness measurements for each star in a Comma Separated Value (\texttt{csv}) file. We also saved general information about each image such as the airmass, exposure time, and background levels.
\end{enumerate}

\subsection{Plotting Light curves} \label{sec:plotting}

\begin{figure*}
\centering
\includegraphics[width=\textwidth]{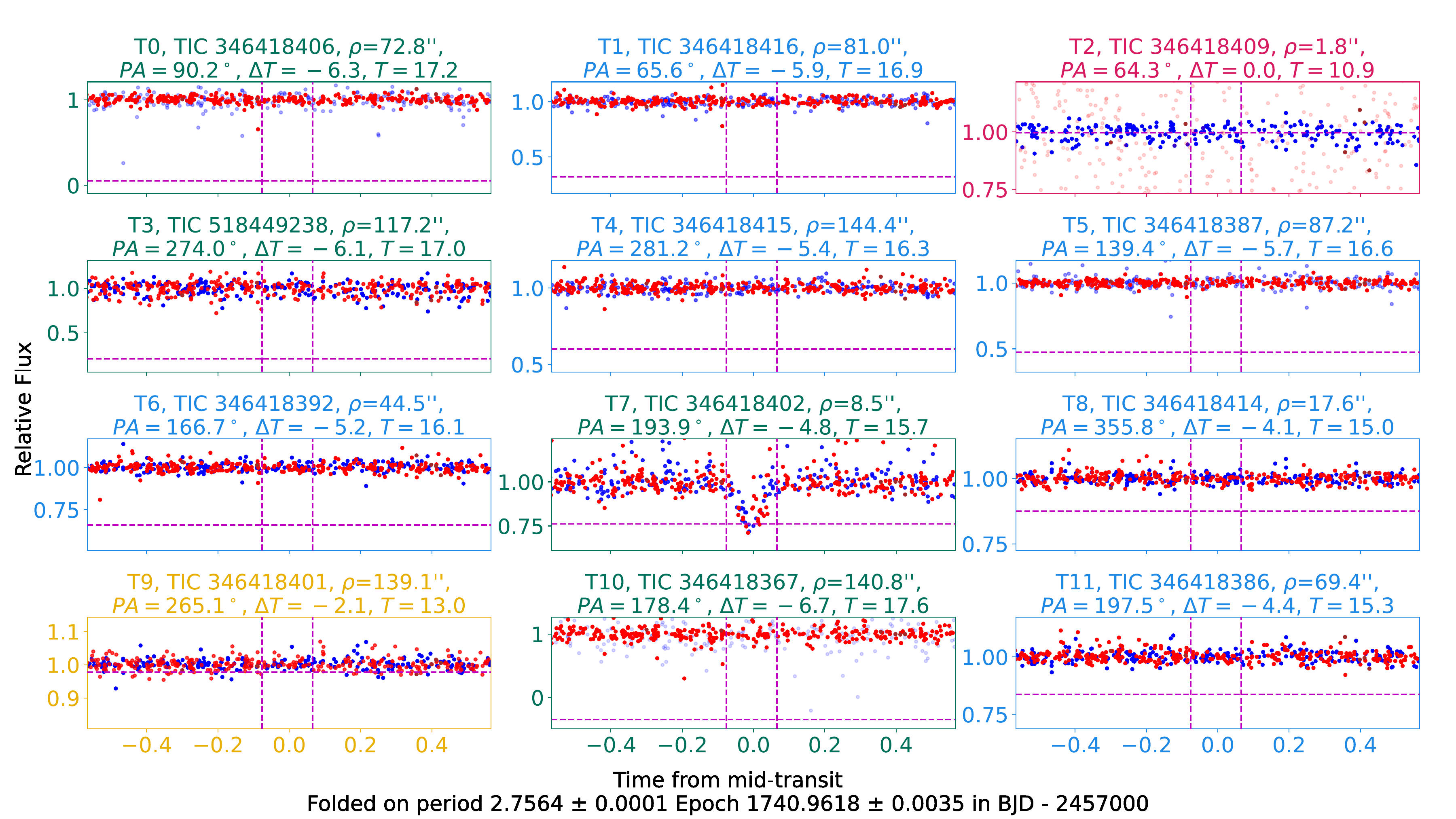}
\caption{ZTF light curves of stars nearby TOI 1423.01 / TIC 346418409. The light curve for the target star, TIC 346418409, is seen in upper-right corner (labeled T2), but the actual source of the signal is TIC 346418402, plotted in the third row from the top, middle column (labeled T7). The color of each subplot border follows the same scheme as the stars in Figure \ref{reference_image} and tells whether the light curve comes from the target star (pink), a comparison star (light blue), a star too saturated to be a comparison star (yellow), a star not used as a comparison star (green), either because it is too faint or because it was manually removed. In this example, despite being bright enough to use as a comparison, the signal star (T7) has astrophysical variability, which could add noise to the other stars' light curves if included as a comparison, so we manually removed it. In each subplot, the vertical dotted magenta lines indicate the start and end times of the transit modulus the period and the horizontal magenta dotted line is how deep the transit of that star would have to be to match the depth measured by \TESS\ in blended light \bedit{(assuming no flux from other sources blends in the \bedittwo{ground-based} aperture)}. Each different color of plotted points corresponds to a different filter from the ground-based telescope data as described in Section \ref{sec:plotting} (blue points show $g$-band, red points show $r$-band, and brown points show $i$-band). The opacity of the points are determined by the standard deviation, i.e. noise level, of all points in that specific filter (greater noise is more transparent). Each subplot is annotated by the star's TIC number, the distance from the target star $\rho$ in arcseconds, the position angle ($PA$) from the target star in degree, the difference in TESS band magnitude between the plotted star and the target ($\Delta T$), and the actual \TESS\ band magnitude $T$. These plots correspond to the stars shown in the reference image in Figure \ref{reference_image}. For more information about this particular example, see Section \ref{sec:TOI 1423.01 / TIC 346418409}. \label{Light Curve}}
\end{figure*}

This section discusses how we produce calibrated light curves from the raw brightness values acquired after taking the steps described in Section \ref{sec:analyzing}. We created custom diagnostic figures which accentuate potential false positives by displaying a light curve for each star in the frame in a subplot. Users can then quickly identify the true transit source among each of the stars within the frame. These  subplots follow the format of example Figure \ref{Light Curve}. To generate these types of diagnostic figures, we take the following steps:

\begin{enumerate}
\item We identified appropriate comparison stars which allow us to normalize the relative brightnesses between frames. To find appropriate comparison stars, we first filtered out stars that had an insufficient number of data points (often stars that were beyond the edge of a particular ZTF field, and were not captured in the same images as the target star). We also required the comparison stars to not be too bright (to avoid artifacts from saturation), to not be too faint (not enough flux to provide useful comparisons to other stars in the field), and to not be photometrically variable, or have other issues that prevent them from being good tracers of systematic effects in the photometry. \bedittwo{On average, we have a range of $10-30$ comparison stars per field (although some very crowded fields can have hundreds). We commonly remove $1$ comparison star manually after our program automatically removes $1-3$ particularly noisy comparison stars.}
\item Next, we identified all of the stars nearby the target that could plausibly be the source of the transit signal detected by \TESS. Not all stars nearby the target are bright enough to contribute the signal. We identified stars that are plausibly bright enough to cause the transit signal observed by TESS by finding all stars within $\Delta m$ magnitudes of the target star, where the maximum difference in magnitudes $\Delta m$ is given by: 
\begin{equation}
\Delta m = \frac{\log\frac{1}{d}}{\log2.512} + 0.5
\end{equation}
where $d$ is the transit depth measured by TESS (converted from ppm to a fraction), and $0.5$ is a buffer to account for possible errors in these quantities\footnote{\bedittwo{In this calculation we made several assumptions: i) we assume that } \bedittwo{all the light of the contaminant is included in the \tess\ aperture. If this is not true, then the actual depth of the eclipsing binary will be larger than what we predict (and therefore easier to detect with ZTF). ii) Another assumption is that the \tess\ bandpass is similar to the bandpass of the ground-based data. All of the bandpasses are relatively close or overlapping in wavelength, which makes this a reasonable assumption, but for strongly chromatic signals, this could change the true depth from our predicted depth.}}. If a star is too saturated or too dim to be an appropriate comparison star, then it is still included in the diagnostic plot, but not used as a comparison star. We include these targets in our subplots since contaminants that do not make good comparison stars can still be the source of the signal, especially in background eclipsing binaries (BEBs) where the actual source of the signal is $5-6$ magnitudes dimmer than the supposed target.
\item We then use the selected comparison stars to produce light curves for each star within the field. For the target star, the light curve was calculated in the following way:
\begin{equation}
\frac{F_T}{\sum_{i=1}^n F_{C_i}}
\end{equation}
Where the brightness of the target star $F_T$ is divided by the sum of all of the brightnesses of the comparison stars $F_{C_i}$. The light curves for comparison stars were calculated at each timestamp based on the following formula:
\begin{equation}
\frac{F_{C_j}}{\sum_{i=1}^n F_{C_i}, i \neq j}
\end{equation}
Where the brightness of the comparison star was divided by the sum of all of the brightnesses of all the other comparison stars (excluding itself).
\item We performed these calculations separately for each image. We then collected stars in each of the telescope's observing filters and divided these brightness values by the median of all brightnesses from that filter. 
\item We then plotted the light curve, folded on the \bedit{orbital period} measured by TESS, with the expected time of transit from TESS centered. In some cases, the orbital period from TESS was not precise enough to yield a clean phase fold\bedittwo{ed light curve. In these cases,} we refined the orbital period using the ZTF or ATLAS observations (see Section \ref{sec:revising}). 
In addition to plotting the phase-folded calibrated light curves, we also estimate and annotate the plots with the expected transit depth on each nearby star that would be required to cause the transit observed by TESS. To compute this, we used the following equations:
\begin{equation}
F_f = 2.512^{T_T - T_F}
\end{equation}
\begin{equation} \label{expecteddepth}
h = \bedittwo{1 -} \frac{(1 - d)(F_f + 1) - 1}{F_f}
\end{equation}
Where $T_F$ and $T_T$ are the nearby faint comparison star's and target stars' \TESS\ band magnitudes respectively, $F_f$ is the ratio of the faint star flux's flux to the target star's flux, and $d$ is the transit depth measured by \TESS. The resultant $h$ is the expected depth of the transit on any nearby star in the field to cause the signal seen by TESS in the blended image (assuming no flux from other sources blends in the \bedittwo{ground-based} aperture). We plot \bedittwo{$1-$} this value as a horizontal magenta line in our figures (as shown in, for example, Figure \ref{Light Curve}).
\end{enumerate}

\begin{figure*}
\centering
\includegraphics[width=\textwidth]{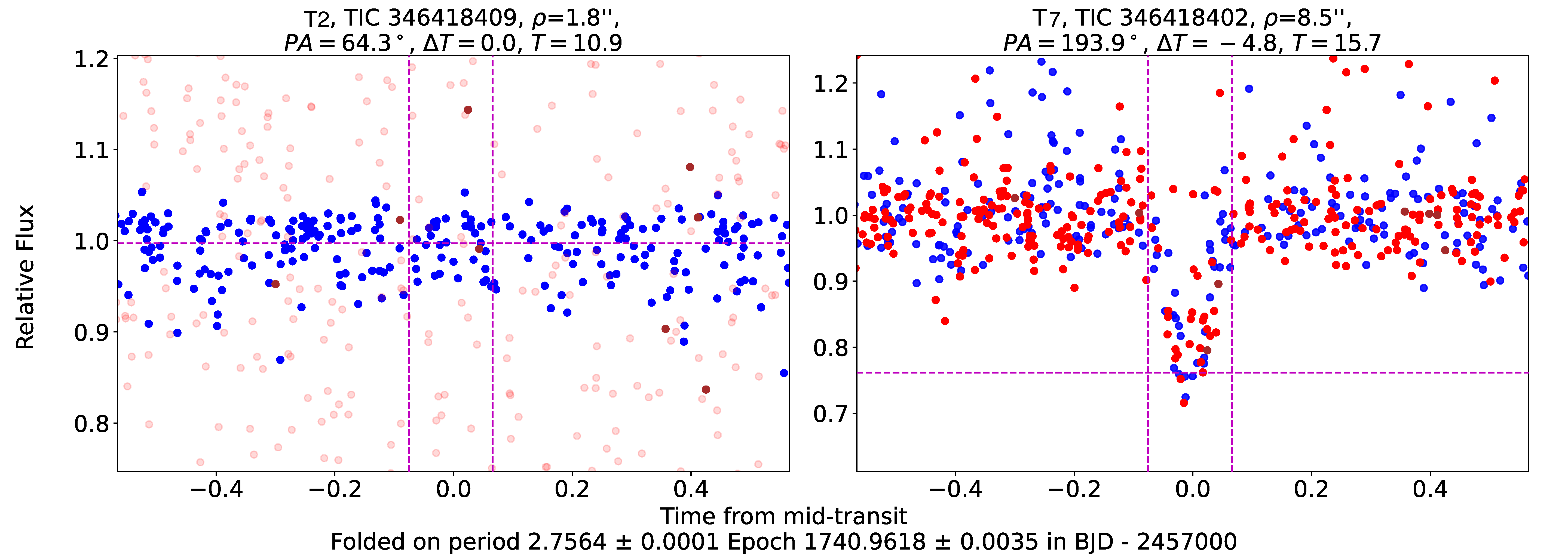}
\caption{ZTF light curves of the suspected planet candidate host star, TOI 1423.01 / TIC 346418409, and the actual source of the transit signal, TIC 346418402. Because the transit is shown from another object, these observations confirm that the planet candidate around the original target is a false positive. The color scheme of the plotted points matches the example full plot in Figure \ref{reference_image}. The upwards outliers in TIC 346418402's light curve are due to its close proximity to the much brigh\bedittwo{t}er target star; images taken in poorer seeing show some blending with the target, contributing additional flux to the aperture. Blending is particularly significant for this object because it is the star with the closest separation between the expected target and the actual false positive source we have confirmed so far. For more details on this target, see Section \ref{sec:TOI 1423.01 / TIC 346418409}. \label{TOI 1423.01 / TIC 346418409}}
\end{figure*}

Figures \ref{reference_image}, \ref{Light Curve}, and \ref{TOI 1423.01 / TIC 346418409} provides a complete set of the reference plots and light curves our program produces. This set of plots is produced for each star within the reference image. The specific example in Figures \ref{reference_image} and \ref{Light Curve} is from TOI 1423.01 / TIC 346418409 using ZTF data.

\subsection{Revising periods and ephemerides} \label{sec:revising}
ZTF and ATLAS data often cover a much longer baseline than the \TESS\ data used to discover the candidates. In many cases, more precise orbital periods for transiting systems can be measured with these ground-based datasets than were possible with \TESS\ alone, and in fact, these more precise orbital periods are sometimes necessary to see the signals in ground-based data. This section discusses how we revise periods after extracting light curves from ground-based surveys. If we do not see a transit at first glance, one possible reason could be that the period from TESS that we use to fold the light curve is slightly incorrect. If the \TESS\-derived period is far enough from the true value, the light curve points taken during transit can be smeared out in phase and blend together outside of the expected time of transit, appearing as uncorrelated noise.

To quantify whether this is a possible cause of a transit non-detection, we calculate the uncertainty on transit times propagated throughout the ground-based dataset, $\sigma_{T_t}$ using:
\begin{equation}
\sigma_{T_t} = \sqrt{(n\sigma_p)^2 + \sigma_{T_0}^2}
\end{equation}
Where $\sigma_{T_t}$ is the uncertainty on the future transit in days, $n$ is the number of transits between the transit epoch time and the reference image time (either oldest or newest image, whichever is larger), $\sigma_p$ is the period uncertainty (provided by \TESS), and $\sigma_{T_0}$ is the transit epoch uncertainty. When this value is much smaller than the transit duration, we expect the TESS ephemeris is sufficiently precise to detect signals without period revision. But when this value is comparable to or larger than the transit duration, the period is too uncertain for our analysis to succeed. 

To address this issue, we use the ZTF and ATLAS data themselves to revise the TESS orbital periods by taking the following steps: 
\begin{enumerate}
\item First, we must identify which star's light curve to use to refine the period. In general, if we do not know the source of a transit, any star in the field could plausibly host the signal, and therefore we would need to perform a period search on each star. This would be time-consuming and prone to false alarm signals, so we usually limit our period searches to either the target (for signals suspected to be true planetary candidates), nearby stars with tentative detections of possible eclipses from other SG1 observations, or nearby stars with suspicious-looking downward outlier points spread throughout the orbital phase that could indicate poorly phased eclipses. 
\item Once we identify which star's light curve to search, we remove as many outlier points as possible. The period search algorithm we use is strongly affected by outliers, so removal of bad points is important for successful period revision. By eye, we set flux levels above and below which we remove all points. We experiment with different levels of outlier exclusion to make sure we do not either remove too few outliers, leaving in too much noise to detect the period, or too many outliers, inadvertantly removing true in-eclipse signals. 
\item After preparing the light curve, we calculate the Box-Least-Squares periodogram (BLS, \citealt{kovacs}) as implemented  in  \texttt{astropy} \citep{astropy:2013, astropy:2018,astropy:2022}. We evaluate that BLS periodogram over a finely-spaced grid of periods ranging from $0.99$ to $1.01$ the original TESS-measured orbital period. 
\item We then plot the resulting BLS periodogram, and by eye, assess whether the periodogram shows a clear peak above the noise in the periodogram. If there is a clear peak, we identify the period of maximum power, which corresponds to the period where the transit dip is the strongest. We then plot the light curve phase-folded on the revised period, and assess whether this improved period yields a clearer detection of a transit/eclipse signal on-target, and if so, record the period for further analysis. 
\item Finally, we then re-run our light curve plotting program using this revised period to examine how it impacts all the stars in the field. 
\end{enumerate}

\begin{figure*}
\centering
\includegraphics[width=\textwidth]{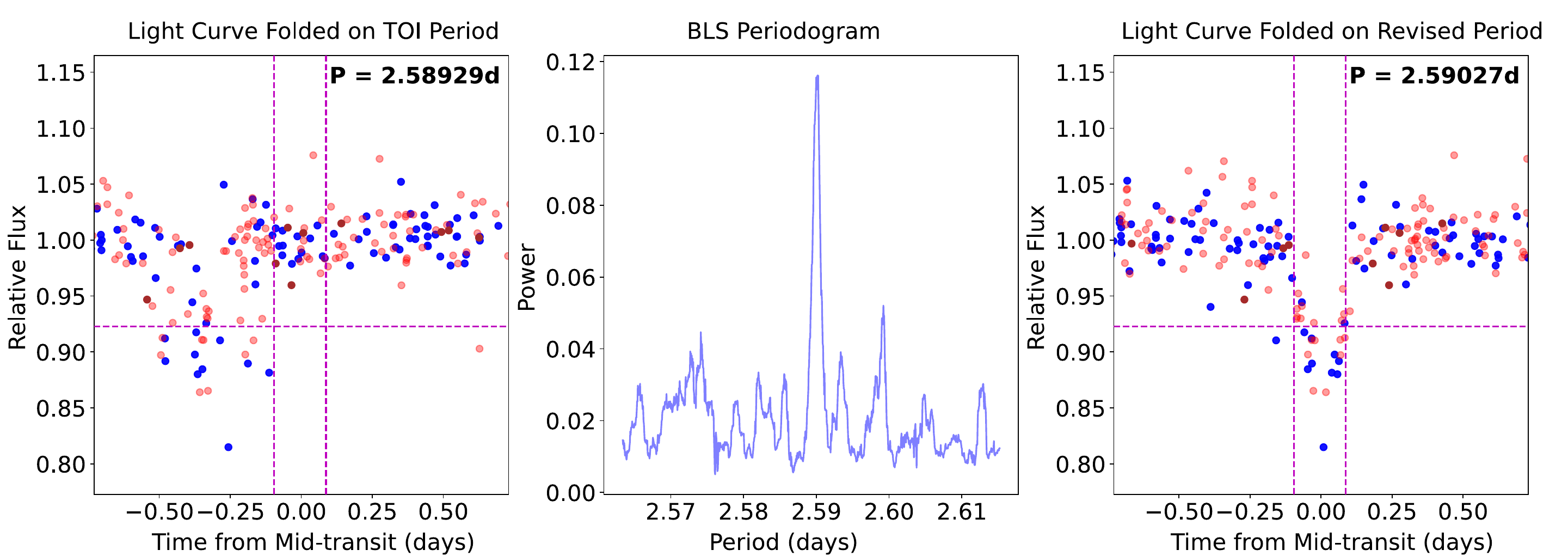}
\caption{ZTF/\deathstar\ light curves of the actual source (TIC 252636890) of the transit signal seen in TOI 1539.01 / TIC 252636888. On the left panel is a transit with a large period uncertainty; propagated over the duration of ZTF observations, the uncertainty on transit times is $0.34$ days, significantly larger than the transit duration.  This uncertainty is large enough to prevent us from seeing a clear dip folded on that period. Nevertheless, we observe a cluster of low points in brightness spread out in the original phase-folded plot, so  so we computed a BLS periodogram to check whether we could revise the period using ZTF. The middle panel shows the BLS periodogram with a clear peak in BLS power at a slightly shorter period of 2.59027d. In the panel on the right, we plot the ZTF lightcurve with the revised period and observe a clear transit at the expected time. For more information about this source, see Section \ref{sec:revising}. \label{Periodogram}}
\end{figure*}

Figure \ref{Periodogram} shows a summary of this process, including a lightcurve phase-folded on the original TESS period before revision, the corresponding BLS periodogram, and the lightcurve after the period revision. \bedittwo{We note that we do not include the TESS data in the period revision yet, but this is a potential avenue for future improvement to our method. }

\section{Results}
\label{sec:Results}
\begin{figure*}
\centering
\includegraphics[width=\textwidth]{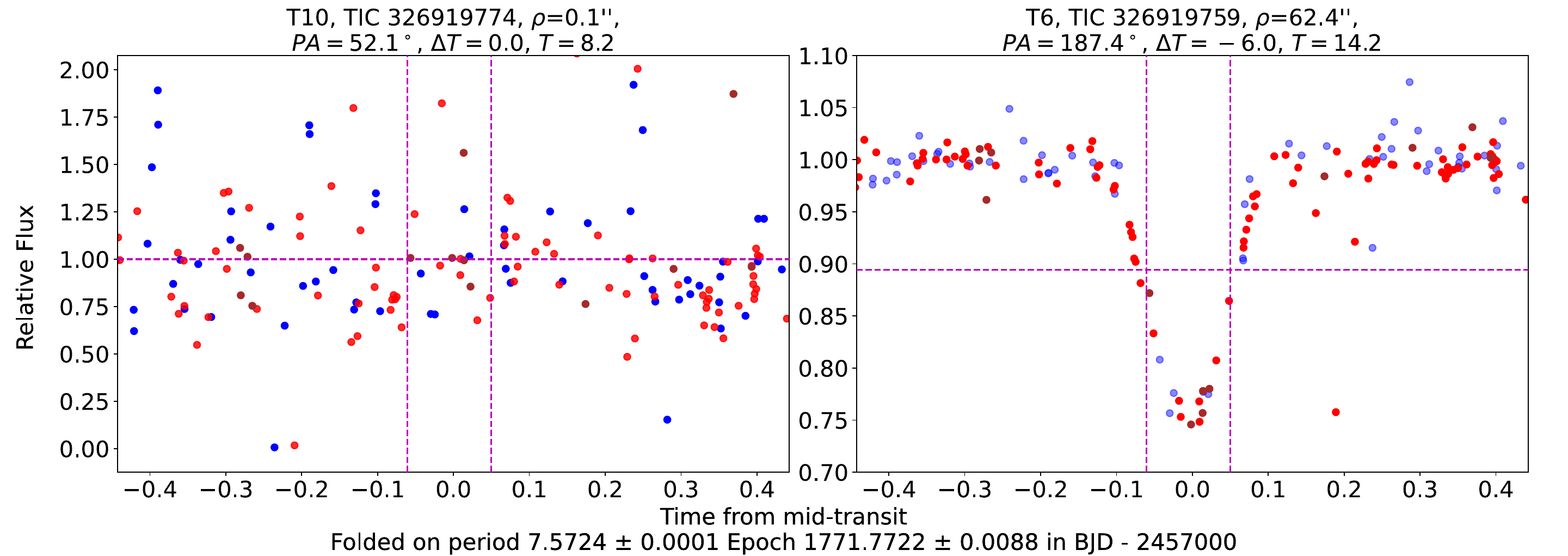}
\caption{ZTF/\deathstar\ light curves of the planet candidate host star TOI 1552 / TIC 326919774 and the nearby star, TIC 326919759, the actual source of the transit signal. Both are folded on the period and epoch of the candidate TOI 1552.01. The clearly visible transit at the predicted orbital period, time of transit, depth, and duration on another star in the field confirms that the original planet candidate is a false positive. The color scheme matches the example full plot in Figure \ref{Light Curve}. This is a particularly clean example of a false positive detection and showcases the high photometric precision achievable with ZTF and \deathstar. We note that the target star itself is highly saturated and thus has a very noisy light curve. \label{TOI 1552.01 / TIC 326919774}}
\end{figure*}

\begin{figure*}
\centering
\includegraphics[width=\textwidth]{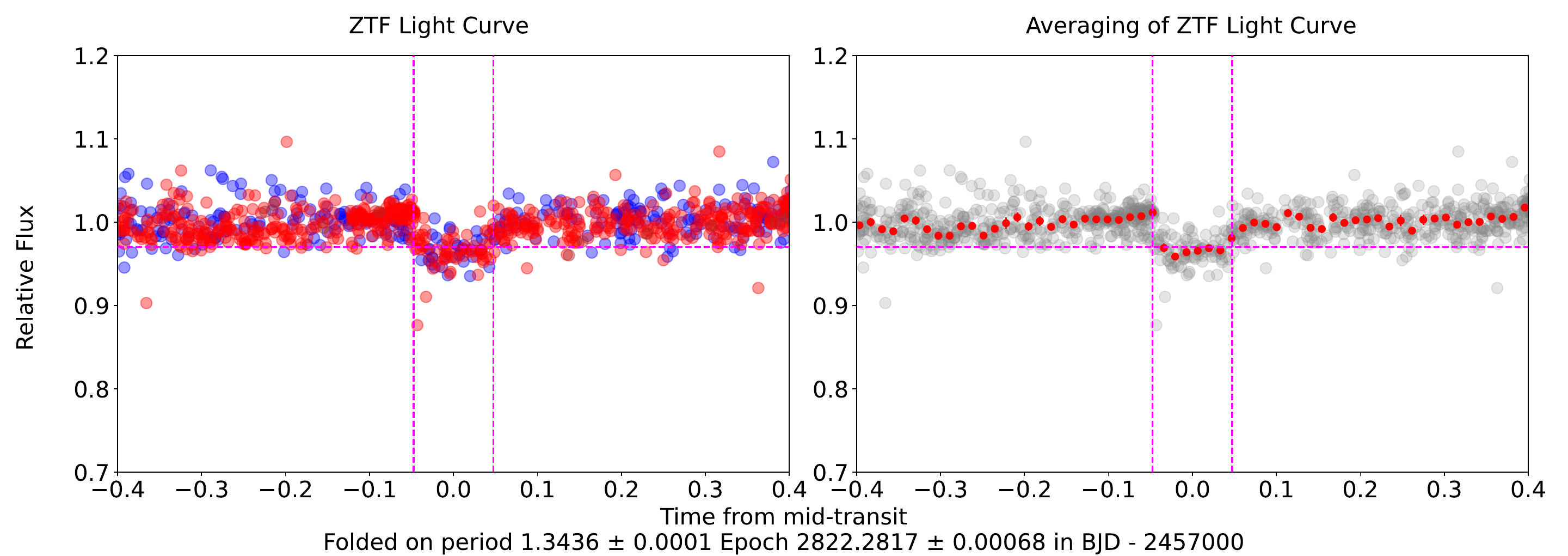}
\caption{ZTF/\deathstar\ light curves of TOI 5871 / TIC 373816781, folded on the period and epoch of candidate TOI 5871.01. The color scheme on the left shows the actual points in their filter colors corresponding to the example full plot in Figure \ref{Light Curve}. The right side of the plot shows the same light curve in grey with average binned points in red with uncertainties. Despite the shallow ($\sim$ 2\%) transit depth, we achieve a confident detection of this signal on-target. \label{TOI 5871.01 / TIC 373816781}}
\end{figure*}

We applied our methodology to a subset of potential planet candidates derived from the list of planet candidates generated from \TESS\ observations. We focused our work on planet candidates that were either already suspected (but not confirmed) to be false positives originating from nearby eclipsing binaries (see, for example, Figure \ref{TOI 1552.01 / TIC 326919774}), or planet candidates with large enough transit depths ($\gtrsim$ 1.5\%) for us to detect \bedit{with ZTF and ATLAS ground-based observations} (see, for example, Figure \ref{TOI 5871.01 / TIC 373816781}). Tables \ref{all_PNEB_table} and \ref{all_VPC_table} summarize our results for false positives and planet candidates respectively confirmed with \deathstar.  In many cases, we were able to confirm the source of transits for signals that were scheduled to be observed by ground-based TFOP resources for confirmation; because we were able to confirm these signals as either false positives or viable planet candidates, we were able to save that telescope observation time for more challenging and promising planet candidates. \bedittwo{So far we have run our system on approximately $100$ planet candidates and confirmed either on-target or off-target signals for $52$: $35$ off-target signals described in Table \ref{all_PNEB_table}, and $17$ on-target signals in Table \ref{all_VPC_table}. In the other cases where we did not confirm a signal, either the light curve was too poor to show a signal or the \tess\ period was too imprecise, or the target had not been observed enough to yield a conclusive result. Our relatively high fraction of conclusive results is because we hand-picked systems where we were likely to detect a signal conclusively (either from previously suspected but unconfirmed false-positives or deeper on-target transits). If we had randomly selected planet candidates to test, we would likely have a smaller fraction of conclusive results.}

Most of our results are based on data from ZTF primarily because of its higher angular resolution and photometric precision, but in some cases, we also incorporated ATLAS observations. ATLAS is particularly helpful in cases where its higher sampling rate is important, such as for long-period signals and for cases where ZTF observes relatively few points. ATLAS will also grow in importance as it collects more observations from the recently added telescopes in the Southern Celestial hemisphere, where ZTF has no observations. 

In the following subsections, we mention five particularly interesting examples of signals successfully confirmed as either being on-target or false positives by \deathstar. Three of these signals demonstrate the extreme performance limits we found with \deathstar: the false positive with the smallest distance from the target (Section \ref{sec:TOI 1423.01 / TIC 346418409}), the longest-period signal (Section \ref{sec:TOI 5478.02 / TIC 95122849}), and shallowest transit (Section \ref{sec:sample_plots}) detected. We also show that \deathstar\ can help reject even high priority (Level 1) \tess\ targets (like sub-Neptune-sized TOI 4148.01 in Section \ref{sec:TOI 4148.01 / TIC 137157546}). Finally, we demonstrate the power of combining ZTF and ATLAS on the same star (Section \ref{sec:TOI 6022.01 / TIC 455947620}). In these sections, we use all stellar and planetary parameter values from  ExoFOP\footnote{\raggedright\url{https://exofop.ipac.caltech.edu/tess/}} \citep{2019AAS...23314009A}.


\subsection{TOI 1423.01 / TIC 346418409} \label{sec:TOI 1423.01 / TIC 346418409}
TOI 1423.01 / TIC 346418409 was originally announced as a planet candidate by the \TESS\ mission team on 6 November 2019\footnote{\raggedright\url{https://tev.mit.edu/data/collection/193/}}, after being identified by the \TESS\ Science Processing Operations Center (SPOC) pipeline \citep{2020RNAAS...4..201C} based out of NASA Ames Research Center. At the time, it was believed to be a candidate super-Neptune-sized ($R_p = 6.41 \pm 7.74$ \rearth) planet in a short-period ($P = 2.75565 \pm 0.00084$ day) orbit around a K3-type star slightly smaller than the Sun ($R_\star = 0.67 \pm 0.043$ \rsun). The full planet candidate parameters are reported online\footnote{\raggedright\url{https://exofop.ipac.caltech.edu/tess/target.php?id=TOI1423}}. Follow-up TFOP SG1 observations, however, did not confirm the transits on-target and instead detected an eclipse at the predicted time of transit on a nearby ($8$\farcs$5$ away) star called TIC 346418402 about $4.8$ magnitudes fainter than the target. The candidate was then given a working disposition by TFOP SG1 of being a ``Potential Nearby Eclipsing Binary'' (PNEB), and was scheduled for follow-up observations with the Las Cumbres Observatory Global Telescope Network to confirm whether the eclipse detected on the nearby star had the same period as the TESS detection, and therefore was its source.

Because of TOI 1423.01's PNEB status, we downloaded and analyzed the ZTF observations of this star using \deathstar. We found in our analysis that the actual source of the signal was indeed the same star detected by the original TFOP SG1 observations, and not the star originally identified by the \TESS\ team as the planet candidate host. We show the \deathstar\ light curves for both the original candidate host star (TOI 1423.01) and the nearby eclipsing binary host (TIC 346418402) in Figure \ref{TOI 1423.01 / TIC 346418409}. The light curves produced by \deathstar\ clearly show $33\%$ deep eclipses in both the ZTF green and red filters -- precisely matching the depth needed to produce the transit observed by TESS (Equation \ref{expecteddepth}). The timing and duration of the eclipses seen in ZTF also match the expectation from \TESS. We consider this as definitive confirmation that TOI 1423.01 is a false positive caused by a nearby eclipsing binary and that the true source of the signal is TIC 346418409.  

Among the signals we have investigated so far, this one is especially interesting because it is the false positive with \bedittwo{one of the closest distances} from the target to the signal source (about 8\farcs5). \bedittwo{As of now, our closest distance for confirming a false positive is about 7\farcs2 on TIC 407394747 (see Table \ref{all_PNEB_table} TOI 1623).} Given the typical angular \bedittwo{resolution} of ZTF, it is difficult to get clean light curves for faint nearby stars closer than about this distance to a bright (often saturated) target star.

\subsection{TOI 5478.02 / TIC 95122849} \label{sec:TOI 5478.02 / TIC 95122849}
TOI 5478.02 / TIC 95122849 was first found by the MIT/TESS Quick Look Pipeline and then released as a planet candidate by the \TESS\ team on 17 May 2022\footnote{\raggedright\url{https://tev.mit.edu/data/collection/193/}}. It was thought to be a hot sub-Saturn-sized planet candidate with a radius roughly $9$ times that of Earth\footnote{The TOI catalog lists a radius of $R_p = 9.15$ \rearth\ with no uncertainty.} and a long period of $10.11970 \pm 0.00015$ days. It was thought to have orbited around a subgiant host star slightly larger than the size of the Sun ($1.748 \pm 0.097$ \rsun). More information on this candidate is reported on ExoFOP\footnote{\raggedright\url{https://exofop.ipac.caltech.edu/tess/target.php?id=TOI5478}}. TFOP SG1 did not confirm the transit signal on-target, and instead detected a potential eclipse at the predicted time of transit and period on a nearby ($10$\arcsec\ away) star called TIC 95122851, $5.2$ TESS magnitudes fainter than the predicted source. The candidate's status was changed to a PNEB and scheduled for follow-up observations to confirm its source.

\begin{figure*}
\centering
\includegraphics[width=\textwidth]{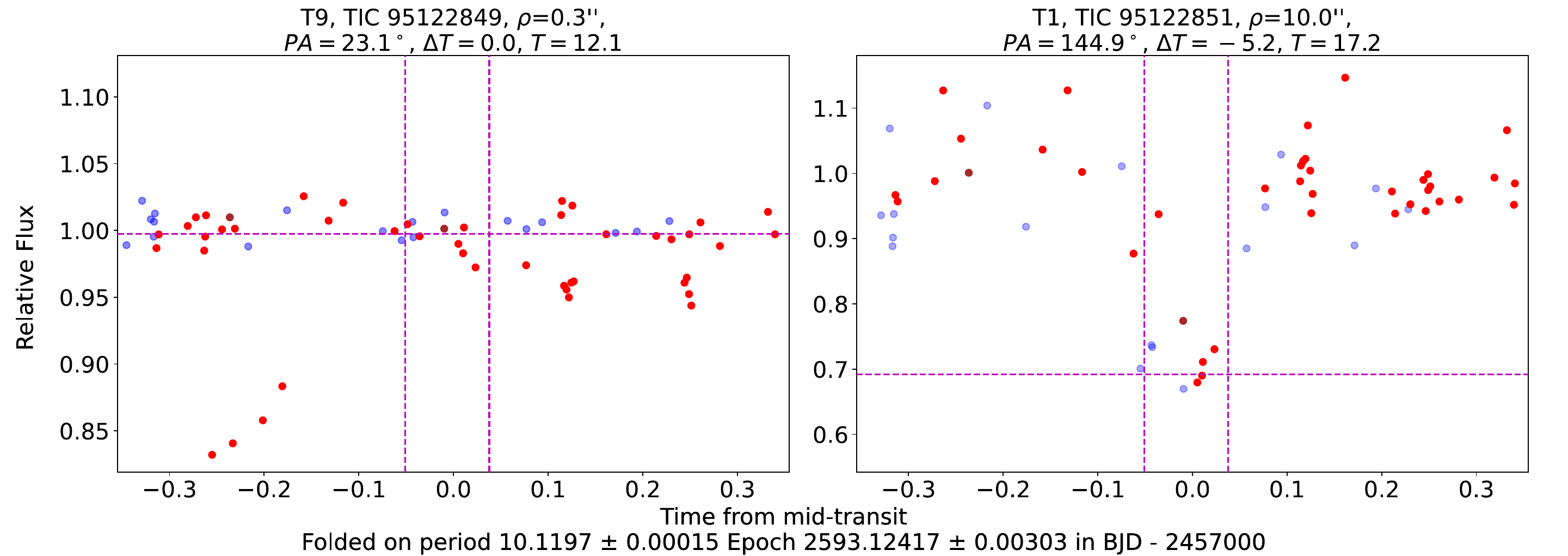}
\caption{ZTF/\deathstar\ light curves of suspected planet candidate host TOI 5478 / TIC 95122849 versus the actual source of the signal, TIC 95122851, folded on the period and epoch for the second candidate TOI 5478.02. The color scheme matches the example full plot in Figure \ref{Light Curve}. The off-target detection confirms that the planet candidate at the original target is a false positive. This target is interesting in that it shows one of our longest periods for which we achieved conclusive results. Long periods are challenging to measure with ground-based data, given gaps in observations due to weather, daytime, and other disturbances. For the full discussion on this target, see Section \ref{sec:TOI 5478.02 / TIC 95122849}. \label{TOI 5478.02 / TIC 95122849}}
\end{figure*}

We downloaded the ZTF telescope images of this PNEB and analyzed them using \deathstar. We found that the actual source was the same star detected by the original TFOP SG1 observations instead of the star initially identified by \TESS. We show the \deathstar\ light curves for both the original candidate host star (TOI 5478.02) and the nearby eclipsing binary host (TIC 95122851) Figure \ref{TOI 5478.02 / TIC 95122849}. These light curves show an eclipse in all three ZTF filters at the period and time predicted by TESS and a depth ($32\%$) precisely matching the depth needed to produce the transit observed by TESS (Equation \ref{expecteddepth}). We consider this a definitive confirmation that TOI 5478.02 is a false positive caused by a nearby eclipsing binary and the actual signal is coming from TIC 95122851. 

This result stands out because the signal has one of the longest periods we investigated at $10.11970 \pm 0.00015$ days. Usually gaps in ground-based observations from the day/night cycle and poor weather make it difficult and thereby rare to detect and confirm periods longer than $4$ days \citep{gaudi2005}. This particular TOI is also interesting because it was part of a multi-planet-candidate system where both candidates are likely false positives. Typically, candidates in multi-candidate systems are very likely to be true planets \citep[e.g.][]{latham2011, lissauer, rowe}, but the TOI 5478 system is an example where one candidate (TOI 5478.01) appears to be an on-target false positive (based on reconnaissance spectra uploaded to ExoFOP) and the other candidate (TOI 5478.02) is confirmed by \deathstar\ to be an off-target false positive.

\subsection{TOI 4198.01 / TIC 459913687} \label{sec:sample_plots}
TOI 4198.01 / TIC 459913687 was announced as a planet candidate by the \TESS\ mission team's QLP Faint Star Search \citep{kunimoto2022} on 12 August 2021\footnote{\raggedright\url{https://tev.mit.edu/data/collection/193/}}. It is believed to be a Jupiter-sized ($R_p = 12.2509$ \rearth with no uncertainty information provided) planet with a short period ($P = 4.74632 \pm 0.00085$ day) around a star slightly larger than the Sun ($R_\star = 1.0907$ \rsun with no uncertainty information provided). A summary of follow-up observations and parameters is available online\footnote{\raggedright\url{https://exofop.ipac.caltech.edu/tess/target.php?id=TOI4198}}. SG1 conducted follow-up observations to try to detect the signal of the planet on-target, but because of the large uncertainty in the candidate's orbital period, targeted ground-based observations were unsuccessful in detecting the signal. 

\begin{figure*}
\centering
\includegraphics[width=\textwidth]{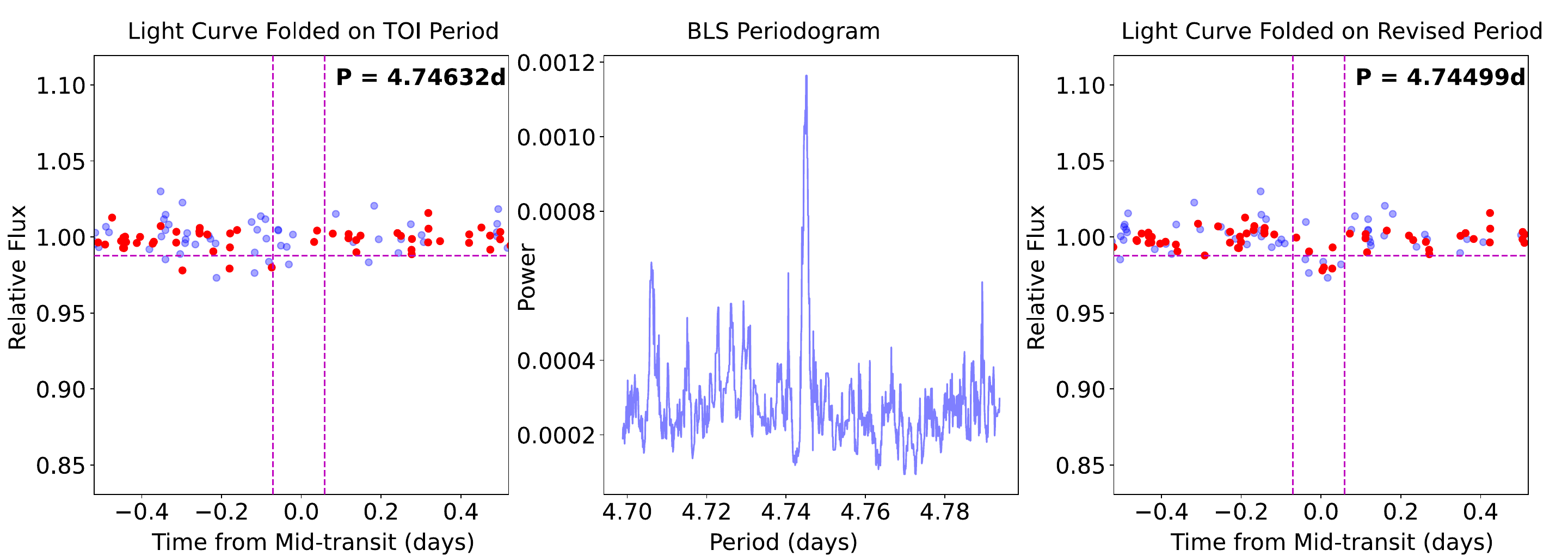}
\caption{ZTF/\deathstar\ light curves of TOI 4198 / TIC 459913687 and the light curve's BLS periodogram. The left-hand panel was folded on the period estimated from \TESS\ observations of the candidate TOI 4198.01, $P= 4.74632 \pm 0.00085$ days, and showed no clear signal. However, a BLS periodogram (middle panel, as described in Section \ref{sec:revising}) revealed a strong peak at a slightly different period, $P= 4.74499$ days. The right-hand panel shows the light curve folded on this new period, and reveals a tentative detection of the signal on-target. The color scheme matches the example full plot in Figure \ref{Light Curve}. This signal, with a depth of about 1.5\%, is the shallowest transit we have detected so far with \deathstar. For the full description of this target, see Section \ref{sec:sample_plots}. \label{TOI 4198.01 / TIC 459913687}}
\end{figure*}

We downloaded and analyzed ZTF data using \deathstar\ to see if we could detect the signal of the candidate and provide a revised orbital period to inform future SG1 observations. Through our analysis we confirmed that the actual source of the signal was on the suspected source of the signal, TIC 459913687. We show the \deathstar\ light curves for the original candidate host star (TOI 4198.01) in Figure \ref{TOI 4198.01 / TIC 459913687}. A BLS search of the \deathstar\ light curve of the target star yielded the detection of a $\approx 2\%$ deep eclipse in ZTF filter bands, consistent with the 1.5\% transit detected by TESS. The detected transit's duration and orbital phase also matched the signal reported by TESS, so we conclude that TOI 4198.01 is an on-target detection.

TOI 4198.01 is the shallowest signal we have been able to detect with \deathstar\ observations. Given the relatively small number of points (compared to ground-based exoplanet surveys like KELT) and the limited precision for \bedit{ZTF} ground-based observations, this object is close to the practical limiting depth for conclusive detections. While most planet transits are shallower than 1.5\%, TESS has identified a large number of candidates (and real planets) with depths greater than this limit. This system also demonstrates the power of \deathstar\ to revise orbital periods. The original period from TESS observations using \tess\ has a period $P = 4.74632 \pm 0.00085$ days, but when we analyzed this detection, we found that the period is actually closer to $P = 4.74499$ days, which is a large enough difference to explain the non-detection by earlier SG1 observations of the candidate. While our revised period still awaits confirmation from additional ground-based resources, we are confident in the detection given the convincing spike in the BLS periodogram (Figure \ref{TOI 4198.01 / TIC 459913687}) and the consistency of the duration and orbital phase with TESS detection.

\subsection{TOI 4148.01 / TIC 137157546} \label{sec:TOI 4148.01 / TIC 137157546}
TOI 4148.01 / TIC 137157546 was originally announced as a planet candidate by the \TESS\ mission team on 23 June 2021\footnote{\raggedright\url{https://tev.mit.edu/data/collection/193/}}, after being identified by the MIT/\TESS\ Quick Look Pipeline (QLP) as part of the Faint Star Search \citep{kunimoto2022}. At the time, it was believed to be a candidate sub-Neptune-sized ($R_p = 3.74 \pm 0.28$ \rearth) planet in a short-period ($P = 3.6841446 \pm 0.0000721$ day) orbit around a K3-type star slightly smaller than the Sun ($R_\star = 0.79 \pm 0.05$ \rsun). The full planet candidate parameters are reported online\footnote{\raggedright\url{https://exofop.ipac.caltech.edu/tess/target.php?id=TOI4148}}. Follow-up TFOP SG1 observations, however, did not confirm the transits on-target and instead detected an eclipse at the predicted time of transit on a nearby ($13$\farcs$6$ away) star called TIC 137157545 about $5$ magnitudes fainter than the target. The candidate was then given a working disposition by TFOP SG1 of a PNEB, and was scheduled for follow-up observations with the Las Cumbres Observatory Global Telescope Network to confirm that the eclipse detected on the nearby star was the source of the transits.

\begin{figure*}
\centering
\includegraphics[width=\textwidth]{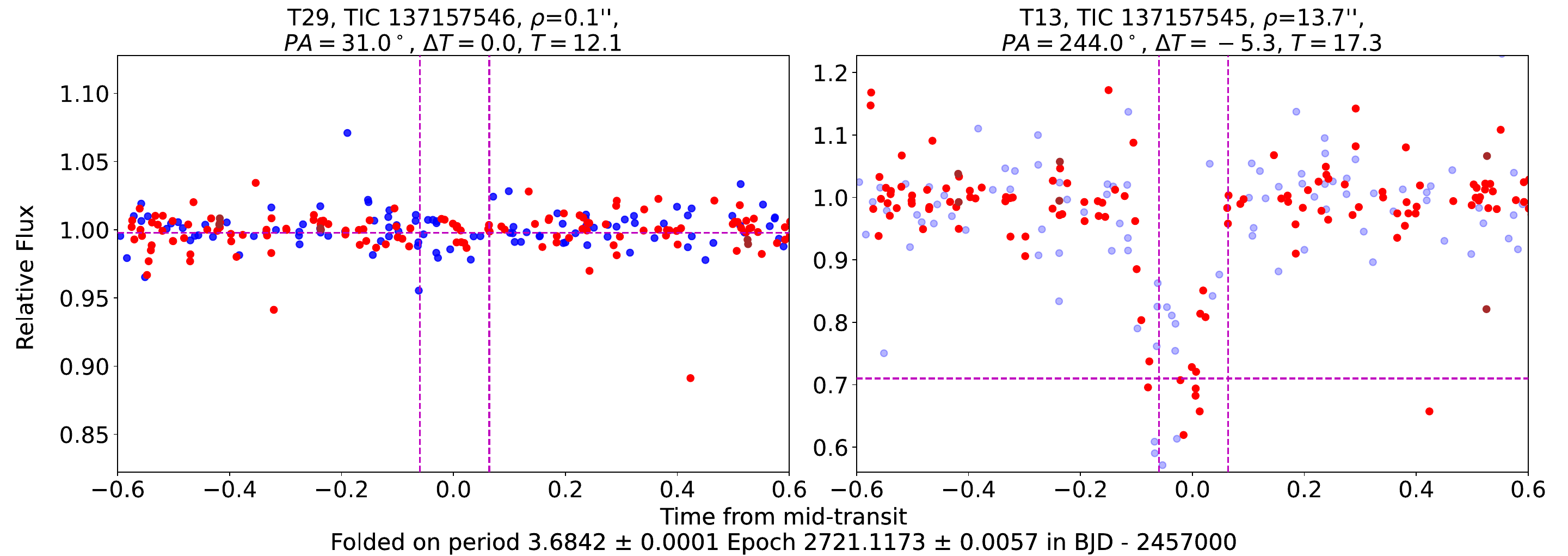}
\caption{ZTF/\deathstar\ Light curve of the suspected planet candidate host star, TOI 4148 / TIC 137157546, and the actual source of the signal, TIC 137157545, both folded on the period and epoch of the candidate TOI 4148.01. Again, the color scheme for the plotted points is the same as described in Figure \ref{Light Curve}. The off-target detection confirms that this is a false positive planet candidate, despite the flat bottomed shape of the transit, often seen as an indication of a high quality planet candidate. Because the candidate was sub-Neptune in size and had a corresponding transit shape, it would have been a higher priority target for more time-intensive observations, like precise radial velocity follow-up. Refuting this tricky planet candidate with archival images prevents other teams from investing expensive telescope observations and saves those resources for real planets. We note that the transit is not perfectly centered at the predicted time, which likely indicates we are folding on a slightly inaccurate orbital period. For more information on this target, see Section \ref{sec:TOI 4148.01 / TIC 137157546}. \label{TOI 4148.01 / TIC 137157546}}
\end{figure*}

Because of TOI 4148.01's status as a PNEB, we downloaded and analyzed the ZTF observations of this star using \deathstar.  We found in our analysis that the actual source of the signal was indeed the same star detected by the original TFOP SG1 observations, and not the star originally identified by the \TESS\ team as the planet candidate host. We show the \deathstar\ light curves for both the original candidate host star (TOI 4148.01) and the nearby eclipsing binary host (TIC 137157545) Figure \ref{TOI 4148.01 / TIC 137157546}. These light curves produced by \deathstar\ clearly show $30\%$ deep eclipses in both the ZTF green and red filters -- precisely matching the depth needed to produce the transit observed by TESS (Equation \ref{expecteddepth}). The timing and duration of the eclipses seen in ZTF also match the expectation from \TESS. We consider this as definitive confirmation that TOI 4148.01 is a false positive caused by a nearby eclipsing binary and that the true source of the signal is TIC 137157545.

This target was our first test of \deathstar's performance on a high priority \TESS\ sub-Neptune-sized planet candidate. A primary goal of the \TESS\ mission was to detect planets smaller than Neptune and measure their masses with ground-based telescopes; this was one of the mission's Level 1 Science Requirements. Due to candidate's small size and the flat bottom of the transit (usually a signature of a likely exoplanet) shown in Figure \ref{TOI 4148.01 / TIC 137157546}, this object might \bedit{have} been confused for a real planet and prioritized for follow-up observations. However, the initial SG1 observations and our confirmation that it is a false positive helped prevent this. 

\subsection{TOI 6022.01 / TIC 455947620} \label{sec:TOI 6022.01 / TIC 455947620}
TOI 6022.01 / TIC 455947620 was reported as a planet candidate by the \TESS\ team on 15 December 2022\footnote{\raggedright\url{https://tev.mit.edu/data/collection/193/}}, after being identified by the \TESS\ Science Processing Operations Center (SPOC) pipeline \citep{Jenkins:2015, Jenkins:2016}, based at NASA Ames Research Center. The transit signal implies a roughly Jupiter-sized planet candidate, with a radius of $13.37 \pm 0.71$ \rearth, orbiting around an M-dwarf star roughly one third the size of the Sun ($R_\star = 0.369 \pm 0.011$ \rsun). The complete information on this TOI is available online\footnote{\raggedright\url{https://exofop.ipac.caltech.edu/tess/target.php?id=TOI6022}}. 

\begin{figure*}
\centering
\includegraphics[width=\textwidth]{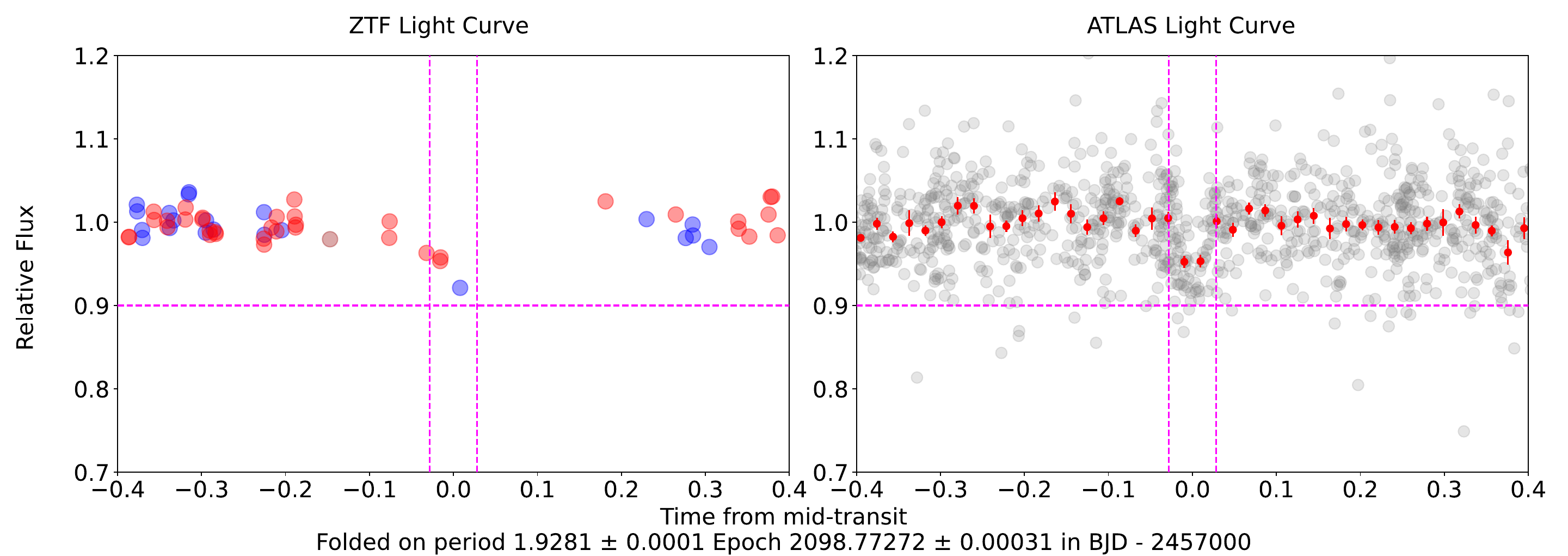}
\caption{\deathstar\ light curves of the planet candidate host star TOI 6022 / TIC 455947620 in both ZTF (left) and ATLAS (right) datasets, folded on the period and epoch of the candidate TOI 6022.01. Both light curves show evidence for an on-target, roughly 8\% deep transit, confirming the suspected planet candidate host is in fact the source of the signal. The ZTF color scheme on the left matches the example full plot in Figure \ref{Light Curve}. On the right, for visual clarity, we show all points as grey dots, regardless of whether they were observed in ATLAS' orange or cyan filters. Bold red points with uncertainties are averages of brightness in bins of orbital phase. Even though ZTF did not have enough points to confirm the end of the transit, ATLAS shows that the on-target detection is certain. Interestingly, the signal we observe is somewhat shallower than predicted by \TESS\ data; this shallower depth has since been confirmed by other ground-based SG1 observations. For a more detailed discussion of this target, see Section \ref{sec:sample_plots}.  \label{TOI 6022.01 / TIC 455947620}}
\end{figure*}

Because this was a large planet candidate orbiting a small star, and therefore has a large transit depth, we chose to download and analyze ZTF observations of TOI 6022.01  to try to detect the transit signal on target. The ZTF light curve was sparsely sampled during the actual transit, but we saw tentative evidence for an on-target signal (three low points during transit). We then decided to download ATLAS data to confirm the detection. ATLAS has worse photometric precision, but many more observations than ZTF for this object, including many during transit. We show the \deathstar\ light curves for the candidate host star (TOI 6022.01) from both ZTF and ATLAS datasets in Figure \ref{TOI 6022.01 / TIC 455947620}. The ATLAS light curve in Figure \ref{TOI 6022.01 / TIC 455947620}b  conclusively shows that the transit signal indeed originates from TIC 455947620. Both ATLAS and ZTF light curves show a $9\%$ deep transit (roughly matching the depth reported by TESS) at the expected time, period, and duration predicted by TESS.

TOI 6022.01 demonstrates the synergistic power of combining data from both ATLAS and ZTF. In this case (and many others), the ZTF light curve is highly precise, but has poor sampling in-transit. ZTF alone hints at an on-target detection, but by itself is unconvincing. On the other hand, the ATLAS data is less precise, but has many more data points than ZTF, which allowed us to confirm this event. ATLAS has lower quality data per point but significantly more observations, such that we can average the data points via binning to get a clear detection. Combining a small number of highly precise points with a larger number of noisier observations from \bedit{multiple} ground-based surveys will be an important strategy going forward and push \deathstar\ to longer orbital periods and more challenging observations. 

\begin{center}
\begin{table*}
\begin{tabular}{@{}|>{\centering\arraybackslash}>{\bfseries}m{1.05cm}|>{\centering\arraybackslash}m{1.3cm}|>{\centering\arraybackslash}m{1.25cm}|>{\centering\arraybackslash}m{1.45cm}|>{\centering\arraybackslash}m{1.35cm}|>{\centering\arraybackslash}m{1.5cm}|>{\centering\arraybackslash}m{1.5cm}|>{\centering\arraybackslash}m{1.25cm}|>{\centering\arraybackslash}m{1.5cm}|>{\centering\arraybackslash}m{1cm}|>{\centering\arraybackslash}m{1cm}|>{\centering\arraybackslash}m{1cm}|>{\centering\arraybackslash}m{1.5cm}|@{}}
\hline
Target TOI & \textbf{Target TIC} & \textbf{Period from TOI Catalog (days)} & \textbf{Period Uncertainty on TOI Period (days)} & \textbf{Period Used if Different (days)} & \textbf{Period Uncertainty on Period Used (days)} & \textbf{Source of Revised Period} & \textbf{Transit Duration (hours)} & \textbf{Transit Epoch (BJD)} & \textbf{Transit Depth from TESS (\%)} & \textbf{Target Tmag} & \textbf{Signal Tmag} & \textbf{Distance Between Target and Signal (arcseconds)} \\
\hline
526.01     & 200593988  & 7.6991                         & 0.0033                                  & 7.69399                         & 0.00048                                  & \deathstar\ & 4.4                      & 2458470.849         & 0.81                         & 12.3        & 15.8        & 17.6                                       \\
\rowcolor{table_alternate}
644.01     & 63303499   & 1.927133                       & 0.00021                                 & 1.927326                        & 0.000108                                 & \deathstar\ & 4.6                      & 2459202.296         & 0.059                        & 10.2        & 16.5        & 11.0                                         \\
971.01     & 177722855  & 2.3904                         & 0.0012                                  & 2.38808                         & 0.00018                                  & \deathstar\ & 6.0                        & 2459203.897         & 0.091                        & 10.5        & 16.0          & 21.3                                       \\
\rowcolor{table_alternate}
1314.01    & 136848581  & 2.63999                        & 0.00043                                 &                                 &                                          & & 2.8                      & 2458713.141         & 0.083                        & 9.9         & 16.3        & 19.3                                       \\
1330.01    & 357457104  & 1.748054                       & 0.00000061                              &                                 &                                          & & 2.1                      & 2459852.149         & 0.067                        & 9.5         & 16.3        & 9.5                                        \\
\rowcolor{table_alternate}
1353.01 & 279177746 & 4.435003 & 0.000010 & & & & 1.9 & 2459850.822 & 0.12 & 10.2 & 15.7 & 24.6 \\
1403.01    & 328750515  & 2.335962                       & 0.000808                                & 2.33712                         & 0.00036                                  & \deathstar\ & 6.0                        & 2459825.345         & 0.31                         & 10.7        & 14.0          & 17.0                                         \\
\rowcolor{table_highlight}
1423.01    & 346418409  & 2.75565                        & 0.00084                                 & 2.75639                         & 0.00012                                  & SG1 & 3.4                      & 2458740.962         & 0.29                         & 10.9        & 15.7        & 8.5                                        \\
1506.01 & 64836837 & 2.26668 & 0.00067 & & & & 2.5 & 2459877.491 & 0.18 & 9.8 & 14.8 & 18.4 \\
\rowcolor{table_alternate}
1508.01    & 415741431  & 2.1568                         & 0.0004                                  & 2.1576                          & 0.000006                                 & SG1 & 2.2                      & 2458739.705         & 0.1                          & 10.5        & 13.7        & 37.1                                       \\
\rowcolor{table_highlight}
1539.01    & 252636888  & 2.590266                       & 0.000002                                & 2.59027                         & 0.00015                                  & \deathstar\ & 4.4                      & 2459880.281         & 0.07                         & 10.4        & 15.5        & 43.6                                       \\
\rowcolor{table_highlight}
1552.01    & 326919774  & 7.5719                         & 0.0028                                  & 7.5724                          & 0.00026                                  & \deathstar\ & 2.7                      & 2458771.772         & 0.042                        & 8.2         & 14.2        & 62.4                                       \\
1593.01    & 192372961  & 2.06313                        & 0.00099                                 & 2.063523                        & 0.000104                                 & \deathstar\ & 2.2                      & 2458792.296         & 0.06                         & 9.1         & 15.5        & 16.5                                       \\
\rowcolor{table_alternate}
1596.01 & 431899140 & 1.97656 & 0.00070 & 1.97693 & 0.00013 & SG1 & 3.1 & 2459928.935 & 0.11 & 10.1 & 12.7 & 15.1 \\
1619.01 & 377144784 & 2.386701 & 0.000010 & 2.38671 & 0.00014 & \deathstar\ & 2.9 & 2459882.667 & 0.057 & 10.1 & 14.5 & 7.2 \\
\rowcolor{table_alternate}
1623.01 & 407394748 & 0.7522768 & 0.0000005 & & & & 1.1 & 2459935.743 & 0.12 & 9.5 & 14.4 & 7.2 \\
2060.01 & 285542903 & 2.26668 & 0.00067 & 2.2658 & 0.00014 & SG1 & 3.0 & 2459883.838 & 0.032 & 9.9 & 16.6 & 16.8 \\
\rowcolor{table_alternate}
2744.01    & 279989567  & 0.787177                       & 0.00000036                              &                                 &                                          & & 1.5                      & 2459226.189         & 0.14                         & 11.5        & 17.4        & 14.5                                       \\
2771.01    & 437893926  & 3.8454384                      & 0.0000015                               &                                 &                                          & & 4.7                      & 2459548.185         & 0.46                         & 12.4        & 14.5        & 17.0                                         \\
\rowcolor{table_alternate}
2838.01    & 81231810   & 5.4521068                      & 0.0000034                               &                                 &                                          & & 4.7                      & 2459223.265         & 0.82                         & 12.9        & 15.1        & 12.2                                       \\
2878.01    & 4999813    & 2.0849517                      & 0.0000022                               &                                 &                                          & & 3.1                      & 2459226.726         & 0.24                         & 12.6        & 15.6        & 18.2                                       \\
\rowcolor{table_alternate}
2936.01    & 22020459   & 1.3328042                      & 0.000001                                &                                 &                                          & & 2.8                      & 2459278.077         & 0.2                          & 12.6        & 15.8        & 12.8                                       \\
3578.01    & 358186451  & 1.3176767                      & 0.00000089                              &                                 &                                          & & 1.7                      & 2459851.521         & 0.053                        & 11.0          & 16.0          & 15.6                                       \\
\rowcolor{table_alternate}
3957.01    & 305550963  & 1.5891628                      & 0.00000062                              &                                 &                                          & & 1.6                      & 2458767.802         & 0.35                         & 13.2        & 17.6        & 8.2                                        \\
3996.01 & 347011522 & 17.20457 & 0.00086 & & & & 5.0 & 2458973.930 & 0.38 & 12.0 & 16.0 & 12.5 \\
\rowcolor{table_alternate}
4091.01    & 289316336  & 3.00387                        & 0.00011                                 &                                 &                                          & & 3.2                      & 2459033.694         & 0.076                        & 11.3        & 17.2        & 18.4                                       \\
\rowcolor{table_highlight}
4148.01    & 137157546  & 3.6841446                      & 0.0000072                               &                          &                                   & & 3.0                        & 2459721.117         & 0.23                         & 12.1        & 17.3        & 13.7                                       \\
\rowcolor{table_alternate}
5418.01    & 115564354  & 1.6421819                      & 0.0000014                               &                                 &                                          & & 2.5                      & 2459550.466         & 0.11                         & 11.7        & 16.8        & 7.3                                        \\
5456.01    & 87044036   & 1.071129                       & 0.000103                                & 1.0712827                       & 0.0000029                                & \deathstar\ & 2.2                      & 2459550.184         & 0.2                          & 12.3        & 16.7        & 19.1                                       \\
\rowcolor{table_highlight}
5478.02    & 95122849   & 10.1197                        & 0.00015                                 &                                 &                                          & & 2.1                      & 2459593.124         & 0.257                        & 12.1        & 17.2        & 10.0                                         \\
5742.01    & 258557847  & 3.965824                       & 0.0000074                               &                                 &                                          & & 2.9                      & 2459765.579         & 0.068                        & 10.0          & 16.3        & 15.8                                       \\
\rowcolor{table_alternate}
5787.01    & 446616059  & 3.0965                         & 0.0033                                  & 3.10706                         & 0.00019                                  & \deathstar\ & 4.7                      & 2459791.415         & 0.051                        & 9.8         & 15.9        & 16.1                                       \\
5910.01 & 273774284 & 17.21827 & 0.00046 & & & & 7.9 & 2459821.923 & 0.48 & 13.4 & 15.6 & 11.1 \\
\rowcolor{table_alternate}
5940.01 & 296898634 & 2.85340 & 0.00076 & 2.85364 & 0.00018 & SG1 & 5.2 & 2459849.326 & 1.58 & 14.2 & 17.4 & 10.1 \\
6023.01    & 341444900  & 9.03296                         & 0.00017                                  &                         &                                   & & 5.9                      & 2459877.877         & 0.016                        & 8.6         & 12.9        & 64.3                                       \\
\hline
\end{tabular}
\caption{This table lists all $35$ planet candidates we identified as false positives by confirming an off-target source using \deathstar. The relative position and \TESS\ band magnitude (Tmag) of the actual signal source are displayed  along with the information for the star originally believed to host the planet candidate. If we used a period other than that listed in the TOI catalog in our analysis, the revised period is shown in conjunction with the original \TESS\ period and its source (including our own period revisions, as described in Section \ref{sec:revising}).  The alternating white/light grey highlight is for visual clarity. The targets \bedittwo{either with figures or} discussed \bedittwo{with} \bedittwo{further} detail in Section \ref{sec:Results} are highlighted in green.\label{all_PNEB_table}}
\end{table*}
\end{center}

\begin{center}
\begin{table*}
\begin{tabular}{@{}|>{\centering\arraybackslash}>{\bfseries}m{1.05cm}|>{\centering\arraybackslash}m{1.3cm}|>{\centering\arraybackslash}m{1.25cm}|>{\centering\arraybackslash}m{1.45cm}|>{\centering\arraybackslash}m{1.35cm}|>{\centering\arraybackslash}m{1.5cm}|>{\centering\arraybackslash}m{1.5cm}|>{\centering\arraybackslash}m{1.25cm}|>{\centering\arraybackslash}m{1.5cm}|>{\centering\arraybackslash}m{1cm}|>{\centering\arraybackslash}m{1cm}|@{}}
\hline
Target TOI & \textbf{Target TIC} & \textbf{Period from TOI Catalog (days)} & \textbf{Period Uncertainty on TOI Period (days)} & \textbf{Period Used if Different (days)} & \textbf{Period Uncertainty on Period Used (days)} & \textbf{Source of Revised Period} & \textbf{Transit Duration (hours)} & \textbf{Transit Epoch (BJD)} & \textbf{Transit Depth from TESS (\%)} & \textbf{Target Tmag} \\
\hline
\rowcolor{table_highlight}
4198.01   & 459913687  & 4.74632                        & 0.00085                                 & 4.74499                         & 0.00018                          & \deathstar\                                  & 3.1                      & 2458458.494         & 1.2                          & 13.4                                            \\
\rowcolor{table_alternate}
5579.01   & 252928337  & 19.8628                        & 0.0035                                  & 19.8727669                      & 0.0005006                                & \deathstar\                                  & 2.0                        & 2459583.261         & 3.1                          & 13.5                                            \\
5833.01 & 213875310 & 2.6722 & 0.0012 & 2.66881 & 0.00013 & \deathstar\ & 2.4 & 2459793.450 & 1.5 & 13.2 \\
\rowcolor{table_alternate}
5836.01   & 354727907  & 8.4498763                      & 0.0000073                               &                                 &                                          &                                   & 1.8                      & 2459786.404         & 7.1                          & 14.5                                            \\
5852.01   & 387844266  & 3.9957                         & 0.0012                                  & 3.99524                         & 0.00011                                  & \deathstar\                                  & 2.1                      & 2459788.338         & 3.8                          & 13.5                                            \\
\rowcolor{table_highlight}
5871.01   & 373816781  & 1.343636                       &                                         &                                 &                                          &                                   & 2.3                      & 2459822.282         & 2.6                          & 12.8                                            \\
5889.01   & 383214426  & 2.40132                        & 0.00018                                 &                                 &                                          &                                   & 2.0                        & 2459823.072         & 2.7                          & 13.0                                            \\
\rowcolor{table_alternate}
5916.01   & 305506996  & 2.36824                        & 0.00066                                 & 2.3670867                       & 0.0000058                                & \deathstar\                                  & 1.9                      & 2459817.69          & 4.9                          & 14.5                                            \\
\rowcolor{table_highlight}
6022.01   & 455947620  & 1.9281621                      & 0.0000036                               & 1.9281027                       & 0.000002                                 & SG1                                  & 1.3                      & 2459098.773         & 11                           & 13.7                                            \\
\rowcolor{table_alternate}
6027.01   & 368678195  & 2.840646                       & 0.000205                                & 2.8405312                       & 0.0000098                                & \deathstar\                                  & 2.7                      & 2459882.348         & 5.8                          & 13.9                                            \\
6034.01   & 388076422  & 2.57626                        & 0.00013                                 & 2.576185                        & 0.000005                                 & SG1                                  & 1.6                      & 2459255.317         & 5.5                          & 13.2                                            \\
\rowcolor{table_alternate}
6035.01   & 323194443  & 3.90739                        & 0.0006                                  & 3.87124                         & 0.00012                                  & \deathstar\                                  & 2.4                      & 2458767.802         & 1.4                          & 13.9                                            \\
6055.01   & 355640518  & 0.8497194                      & 0.000003                                & 0.8496822                       & 0.0000014                                & CTOI                                  & 1.4                      & 2458817.088         & 6.9                          & 14.5                                       \\
\rowcolor{table_alternate}
6101.01   & 19342878  & 1.332825                      & 0.000014                                & 1.333                       & 0.055                                & \cite{2023arXiv230206724M}                                  & 2.0                      & 2458492.005         & 1.9                          & 10.8                                       \\
6207.01   & 432422552  & 9.77124                      & 0.00045                                &                        &                                 &                                   & 4.0                      & 2459881.332         & 3.4                          & 12.1                                       \\
\rowcolor{table_alternate}
6133.01   & 129107501  & 3.65343                      & 0.00029                                & 3.653542                      & 0.000070                                &  \deathstar\                                 & 3.5                      & 2459850.634         & 2.5                          & 11.5                                       \\
6227.01   & 53874375  & 3.09225                      & 0.00046                                & 3.09225                       & 0.00028                                & \deathstar\                                  & 1.9                      & 2459879.249         & 4.4                          & 14.3                                       \\
\hline
\end{tabular}
\caption{This table shows all of the $17$ on-target transit signals that we have confirmed using \deathstar\, listed in TOI number increasing order. If we used a period other than that listed in the TOI catalog, we list the revised period and its source, whether that be from our own \deathstar\ analysis, community uploads to Exo-FOP as a Community TOI (CTOI), or revised timing from prior SG1 observations.  In particular, we confirmed one TOI (6055.01) using a CTOI period and ephemeris uploaded by Luke Bouma from the CDIPS pipeline \citep{bouma} . The alternating light grey highlight is for easier scanning. The targets discussed in detail in Section \ref{sec:Results} are highlighted in green.\label{all_VPC_table}} 
\end{table*}
\end{center}

\section{Discussion}
\label{sec:discussion}
\subsection{Impact of the work}
\bedit{In this paper, we have demonstrated that we can confirm transiting planet candidate false positives as well as on-target planet signals using ground-based light curves from archival surveys like ZTF and ATLAS using our new software tool, \deathstar.} We have performed analyses and yielded definitive results on about three dozen systems so far, which have already saved time by reducing the need for specially scheduled observations to assess these systems. Assuming each object we have listed would require a roughly $5$ hour observation with a ground-based telescope (a $2$-$3$ hour transit plus out-of-transit baseline on either side), our analysis replaced approximately $150$ hours of specially scheduled telescope time by making use of archival data.

So far, \deathstar\ has been more successful at identifying false positives and disproving the existence of planet candidates than confirming planet candidates themselves (hence the origin of the pipeline\bedit{'}s name as \deathstar). This may appear to be contrary to the goal of any exoplanet hunter, which is of course to find a clear transit dip from the actual source of the signal and contribute to the discovery of real planets orbiting stars outside the solar system. However, our work is still valuable because by eliminating false positives without requiring new observations, \deathstar\ helps save observational and funding resources, which can now be used instead to confirm better planet candidates. 

\bedittwo{We note that \deathstar\ can be straightforwardly modified to use data from other surveys, and to screen planet candidates detected by future missions. We have structured our pipeline so that it should be easy to simply swap out the image source from ZTF or ATLAS to some other survey, without modifying the code used to perform the photometric analysis and data visualization. As new surveys like ARGUS \citep{2022PASP..134c5003L}, ATLAS-Teide \citep{2023arXiv230207954L}, and the Legacy Survey of Space and Time (LSST) from Vera Rubin Observatory \citep{2019ApJ...873..111I} come online, we should be able to take advantage of their data products for screening planet candidates. We also note that \deathstar\ could be useful for vetting planet candidates from the upcoming PLATO mission \citep{2014ExA....38..249R}. PLATO will have a similar pixel scale to TESS, and therefore will also suffer from false positive contamination from nearby eclipsing binaries. A future version of \deathstar\ could use data from all of these surveys to confirm or refute PLATO planet candidates.}

\subsection{Prospects for Improving Photometric Precision}
One way we can increase \deathstar's ability to detect on-target signals is by improving the photometric precision of the measurements. While ground-based observations will have inherent limits to their photometric precision, more sophisticated analysis procedures may yield some improvements and make it possible to more reliably detect shallow ($\lesssim$ 2\%) signals on-target. One possible improvement would be in changing our method for centering apertures in each frame. For ATLAS, we rely on the astrometric solution performed by the survey's pipeline, and for ZTF we perform a fit to a single isolated star in the field to measure offsets between the survey pipeline and the true location of stars. Better selection of the star for measuring the offset, or performing our own astrometric solution with multiple stars in the field may yield better placed apertures and more precise light curves. We may also be able to optimize light curves by using different aperture sizes for different stars. Currently, we use a one-size-fits-all aperture size for the stars in the field, but generally faint stars perform best with small apertures designed to minimize background noise, and bright stars perform best with somewhat larger apertures that capture all of the star's flux. Determining a prescription for aperture size as a function of star magnitude could yield improved photometric precision. Finally, incorporating some sort of systematics correction to the light curves could be helpful. Virtually all ground-based transit surveys use some sort of systematics correction, like Trend Filtering \citep{tfa}, External Parameter Decorrelation \citep{epd}, or SysRem \citep{sysrem}. In particular, we suspect that detrending against external parameters like the image full width at half maximum, background levels, or airmass of the observation could yield improvements in at least some of the light curves we observe. 

If we are successful in improving the photometric precision of \deathstar's light curves, we anticipate being able to more confidently confirm situations where we currently cannot detect signals, such as shallow on-target transits and perform more accurate period revision.

\subsection{Prospects for Automation, Scaling Up\bedittwo{, and Drawing Conclusions in the Absence of Clear Detections}}
So far, we have only run our software on a small number of planet candidates (fewer than $100$). Our eventual goal is to run \deathstar\ on the majority of TOIs ($\sim$ $6000$ targets) in an automated fashion. We expect that this should be computationally feasible. Using either ZTF or ATLAS observational data, downloading and analyzing roughly $800$ images to analyze $40$ stars within the frames' fields in processing a single target took approximately $20$ minutes using single-core processing on a standard personal computer. While $20$ minutes can add up quickly when processing $6000$ targets on a single computer core, we can employ multiprocessing, a more-powerful workstation, or use supercomputing resources to generate results significantly faster. Storage requirements for \deathstar\ processing are also minimal. While the image datasets are large (thousands of images per target with sizes of roughly $300$ KB per ZTF fits image, and $600$ KB per ATLAS fits image which features a larger field of view), the images may be deleted after processing, retaining only the much smaller processed light curves. In principle, it should be possible to substantially scale up our \deathstar\ operations to analyze many more TOIs, saving even more telescope time and funding resources.

Currently, however, the limitation of \deathstar\ is not computing power, but human resources: each TOI must be manually selected and requires troubleshooting. Therefore, the main thrust of our future work on \deathstar\ will be to implement computational automation. For example, instead of hand-selecting promising planet candidates for \deathstar\ analysis and setting the code running on each star manually, with sufficient automation, we could simply loop through the entire TOI list. Another potential way to increase the speed of our analysis will be to automatically detect and highlight likely transits in the \deathstar\ light curves. Our current process involves manually viewing each of up to hundreds of light curves of nearby stars to each planet candidate to determine the actual transit source. In future versions of \deathstar, we can automatically fit the light curve of each star to a transit model and subsequently determine which plot contains a possible transit dip consistent with the \TESS\ detection. This will greatly increase the speed with which we can locate the transit source. Beyond this, it would be highly useful to have an automated method for determining when the TOI period needs revision and searching for  the uncertainty of the period to help troubleshoot in cases where a transit is not found. When the period uncertainty is high, the transit may not appear in our plots and alerts us that the period most likely must be corrected to find potential transits. A future automated solution could take this into account and either alert users to the fact that the period is likely off, or better, find the real period and reassess the transit. \bedittwo{Also, future versions of the code could incorporate the TESS data themselves into the period revision process, potentially improving the period measurements beyond what is currently possible with ground-based data alone.}

A final, major improvement to automation involves acquiring and processing the ATLAS data. For most of our confirmations, we used ZTF both because of its high precision and the \bedittwo{convenient} API for acquiring the observations. However, ZTF only observed the sky north of $-30$ degrees and misses the numerous Southern \tess\ planet candidates. This is where ATLAS has a big role to play; its observations cover the entire sky, including the Southern hemisphere thanks to its array of telescopes around the world. ATLAS may have lower resolution in their images and lower precision in their light curves, but it makes up for these limitations by collecting many more points. \bedittwo{Previously, we downloaded ATLAS data manually, but it is now possible to download the images automatically.} Developing automated tools for acquiring ATLAS data will enable us to extend our analyses to the Southern hemisphere and take advantage of its sampling in cases where ZTF observations cannot yield a conclusive result alone. 

\bedit{We also note that ZTF and ATLAS data may be useful for confirming weak signals detected by \TESS\ around faint stars. In many cases, space-based observations from TESS are more precise than ground-based observations (like ZTF and ATLAS), but that is not necessarily true for the faint stars. \TESS's $10$cm lenses provide significantly less collecting area than ZTF and ATLAS, with their $1.2$m and $0.5$m diameter apertures, respectively. As a result, \TESS's photometric precision will be lower than ZTF and ATLAS when photon noise dominates. In these cases, the addition of data from ZTF and ATLAS could contribute significantly to confirming \TESS\ discoveries around faint stars.}

\bedittwo{Finally, in the future, we plan upgrades to \deathstar\ to make it possible to draw conclusions about whether any given signal is a planet, even when we do not conclusively detect a transit signal in the ground-based data. Most planet candidates detected by TESS have transit depths smaller than \deathstar\ can reliably detect from the ground, so in many cases, the absence of detection of the signal off-target is actually a good indicator that the candidate may be a real planet. If none of the other stars in the field have signals with the required transit depth needed to match the TESS observations, then we can conclude that none of the other stars could be the source of the signal. However, conclusively ruling out signals off-target, called ``clearing'' targets of nearby false positives, is more challenging than affirmatively detecting on-target and off-target signals. In particular, two improvements will be important for clearing targets:} 
\begin{enumerate}
\item \bedittwo{\textit{Accounting for blending between stars}: If two (or more) stars have overlapping apertures in ZTF or ATLAS data, we need to take this into account when calculating the expected depth of the transit seen by ZTF and ATLAS for that particular star. Currently, we calculate the depth of the transits in each nearby star assuming each star is well-separated from its neighbors and isolated in the ZTF/ATLAS images (see Equation \ref{expecteddepth}), but if two stars are blended in ZTF or ATLAS, the actual depth we measure will be smaller. Since we compare the actual depth of signals we detect to the expected depth of the transit to determine whether any given star could be the source of the TESS planet candidate signal, having inaccurate (and in particular underestimated) depths could cause problems. If we see a signal, but it appears significantly too shallow to cause the TESS detection, we could conclude that it is just a coincidental alignment of another binary star, or if we don't see a signal because the true depth is shallower than our typical 1\% photometric, we could incorrectly rule out contamination. We therefore must modify our calculation of expected transit depth to account for blended flux in ZTF/ATLAS apertures when applicable.}
\item \bedittwo{\textit{Accounting for uncertainties in the orbital period}: Often the long baseline of ATLAS/ZTF compared to TESS means that the orbital period uncertainty in TESS will be large enough that the transit may be significantly offset or smeared in ATLAS/ZTF. When we see some evidence for a signal, we can often improve the period precision using BLS, but when we detect no signals, it is unclear whether the nondetection is due to an erroneous orbital period or because there are truly no nearby eclipsing binaries. We therefore will need to build automatic methods for assessing whether slight changes in orbital period would change our conclusion that eclipses are ruled out on nearby stars (for example by shifting transit times to gaps in data).}
\end{enumerate}

\bedittwo{We have already cleared some planet candidates (Hord et al. \textit{submitted}; Capistrant/Soares-Furtado et al. \textit{in press}), where we performed manual checks to ensure no issues as described in the previous paragraph, but automation will be required to significantly scale up and more confidently confirm and clear fields with \deathstar.}

\section{Conclusions}\label{sec:conclusions}
We have presented \deathstar, a system for using archival ground-based time-series observations of TESS planet candidate host stars to determine whether the suspected candidate is actually the source of the transit. We have developed software to download images from ZTF and ATLAS, extract light curves, and produce diagnostic plots that make it straightforward to either confirm or refute planet candidates. We have applied our system to dozens of TESS planet candidates and confirmed both on-target transit signals and off-target false positives. This work contributes to the TESS mission by identifying false positives without any additional telescope observations (freeing up time for observations of other, higher-priority candidates) and it contributes to the field of exoplanets by increasing the purity of our planet candidate lists, helping to enable better population studies. 


\section*{Acknowledgements}

\bedittwo{We thank the anonymous referee for their constructive and helpful feedback, which greatly improved the clarity of this manuscript.} ZLD \bedittwo{would like to thank the generous support of the MIT Presidential Fellowship, the MIT Collamore-Rogers Fellowship and} would like to acknowledge that this material is based upon work supported by the National Science Foundation Graduate Research Fellowship under Grant No. 1745302. \bedit{KAC acknowledges support from the TESS mission via subaward s3449 from MIT.} We acknowledge the use of public TOI Release data from pipelines at the TESS Science Office and at the TESS Science Processing Operations Center. Funding for the TESS mission is provided by NASA’s Science Mission directorate. This work is partially based on observations obtained with the Samuel Oschin Telescope $48$-inch and the $60$-inch Telescope at the Palomar Observatory as part of the Zwicky Transient Facility project. ZTF is supported by the National Science Foundation under Grant No. AST-2034437 and a collaboration including Caltech, IPAC, the Weizmann Institute for Science, the Oskar Klein Center at Stockholm University, the University of Maryland, Deutsches Elektronen-Synchrotron and Humboldt University, the TANGO Consortium of Taiwan, the University of Wisconsin at Milwaukee, Trinity College Dublin, Lawrence Livermore National Laboratories, and IN2P3, France. Operations are conducted by COO, IPAC, and UW. The \texttt{ztfquery} code was funded by the European Research Council (ERC) under the European Union's Horizon 2020 research and innovation programme (grant agreement n°$759194$ - USNAC, PI: Rigault). This research made use of Photutils, an Astropy package for detection and photometry of astronomical sources \citep{photutils}. This research has made use of NASA's Astrophysics Data System and the NASA Exoplanet Archive, which is operated by the California Institute of Technology, under contract with the National Aeronautics and Space Administration under the Exoplanet Exploration Program.

\bedit{This work has made use of data from the Asteroid Terrestrial-impact Last Alert System (ATLAS) project. The Asteroid Terrestrial-impact Last Alert System (ATLAS) project is primarily funded to search for near earth asteroids through NASA grants NN12AR55G, 80NSSC18K0284, and 80NSSC18K1575; byproducts of the NEO search include images and catalogs from the survey area. This work was partially funded by Kepler/K2 grant J1944/80NSSC19K0112 and HST GO-15889, and STFC grants ST/T000198/1 and ST/S006109/1. The ATLAS science products have been made possible through the contributions of the University of Hawaii Institute for Astronomy, the Queen’s University Belfast, the Space Telescope Science Institute, the South African Astronomical Observatory, and The Millennium Institute of Astrophysics (MAS), Chile.}

Datasets:
ZTF
ATLAS
MAST
Exo-FOP

Modules:
python \citep{van1995python}: requests, json, math, os, sys, urllib.parse, pickle, datetime, time, calendar, pathlib, pdb, fpdf
astropy: fits
numpy \citep{harris2020array}
pandas \citep{reback2020pandas}, \citep{mckinney-proc-scipy-2010}
scipy \citep{2020SciPy-NMeth}
matplotlib \citep{Hunter:2007}: \text{mpl\_toolkits.axes\_grid1}
ztfquery
photutils.aperture \citep{larry_bradley_2022_6825092}
mpfit
IPython.display \citep{PER-GRA:2007}
PyAstronomy \citep{pya}


\section*{Data Availability}
We have made all of our code available on GitHub for the greater community to branch off of and contribute to more exoplanet discovery with this base pipeline system\footnote{\raggedright\url{https://github.com/vanderburgers/deathstar}}. The ZTF\footnote{\raggedright\url{https://irsa.ipac.caltech.edu/Missions/ztf.html}} and ATLAS\footnote{\raggedright\url{https://fallingstar-data.com/forcedphot/}} images are available for download online.


\bibliographystyle{mnras}
\bibliography{bib} 


\appendix


\bsp	
\label{lastpage}
\end{document}